\documentstyle[12pt,epsf]{article}

\newcommand{\C}{\bf{C}}
\newcommand{\R}{\bf{R}}

\newcommand{\bz}{\bar{z}}
\newcommand{\hz}{\hat{z}}
\newcommand{\hbz}{\hat{\bar{z}}}
\newcommand{\bw}{\bar{w}}
\newcommand{\hN}{\hat{N}}
\newcommand{\hn}{\hat{n}}

\renewcommand{\a}{\alpha}
\renewcommand{\H}{{\cal H}}
\newcommand{\bea}{\begin{eqnarray}}
\newcommand{\ena}{\end{eqnarray}}
\newcommand{\EQ}{\begin{equation}}
\newcommand{\EN}{\end{equation}}
\newcommand{\vs}[1]{\vspace{#1 mm}}

\newcommand{\pa}{\partial}
\newcommand{\nn}{\nonumber\\}
\newcommand{\lan}{\langle}
\newcommand{\ran}{\rangle}

\makeatletter
 
 \@addtoreset{equation}{section}
\makeatother

\begin{document}


\topmargin 0pt
\oddsidemargin 0mm
\renewcommand{\thefootnote}{\fnsymbol{footnote}}

\begin{titlepage}

\setcounter{page}{0}
\begin{flushright}
KEK-TH 716\\
hep-th/0010006\\
\end{flushright}

\vs{15}
\begin{center}
{\Large\bf  Topological Charge of $U(1)$ Instantons} \\
\vs{5}
{\Large\bf on Noncommutative ${\R}^4$}\\
\vs{20}
{\large
Furuuchi \ Kazuyuki}

\vs{10}
{\em Laboratory for Particle and Nuclear Physics,}\vs{1}\\
{\em High Energy Accelerator Research Organization (KEK), }\vs{1}\\
{\em Tsukuba, Ibaraki 305-0801, Japan} \vs{2}\\
{\em Fax : (81)298-64-5755} \vs{1}\\
{\em E-mail: furuuchi@ccthmail.kek.jp}
\end{center}
\vs{10}
\renewcommand{\thefootnote}{\fnsymbol{footnote}}
\setcounter{footnote}{0}
\centerline{{\large\bf{  Abstract }}\footnote[2]{%
Lecture notes based on talks given at
APCTP-Yonsei Summer Workshop on
Noncommutative Field Theories (Seoul),
August 29, 2000, 
Tokyo University (Hongo),
July 17, 2000
and Tokyo Institute of Technology,
September 29, 2000.}}\vs{5}%
\noindent %
Non-singular instantons are shown to exist
on noncommutative ${\R}^4$ even with a 
$U(1)$ gauge group.
Their existence is primarily due to the
noncommutativity of the space.
The relation between $U(1)$ instantons
on noncommutative ${\R}^4$ and the
projection operators acting on the
representation space of the noncommutative
coordinates is reviewed.
The integer number of instantons 
on the noncommutative ${\R}^4$
can be understood as the winding number
of the $U(1)$ gauge field as well as
the dimension of the projection on
the representation space.

\end{titlepage}
\newpage

\renewcommand{\thefootnote}{\arabic{footnote}}
\setcounter{footnote}{0}

\tableofcontents 
\newpage

\section{Introduction}

The concept of smooth space-time
manifold must be modified at the 
Plunk scale due to the quantum fluctuations,
and quantum gravity must explain such space-time
far from being smooth.
Quantum field theory on noncommutative
space is expected to provide a model for the short scale 
structure of quantum gravity, since it has
non-locality which becomes relevant 
at the scale introduced by the noncommutativity.
The discovery of noncommutative field theory 
in certain limit of string theory \cite{CDS} also gives 
concrete motivations for this subject.

Solitons and instantons play important roles in the
nonperturbative analysis of quantum field 
theories.\footnote{For the role of solitons and
instantons on ordinary 
commutative space, see for example
\cite{Cole}\cite{Raja}.}
Noncommutativity of space introduces new ingredients
to the short scale behaviour of the classical solitons.
For example,
in their pioneering work \cite{NS} Nekrasov and
Schwarz showed that
on noncommutative ${\R}^4$
non-singular instantons can exist
even when the gauge group is $U(1)$.\footnote{%
The field strength of the $U(1)$ one-instanton 
solution given in
\cite{NS} must be taken as field strength
in the reduced Fock space \cite{mine}.
This point is also described in the
latter half of the lectures.
}
It is a typical appearance of the
effect of noncommutativity
because in ordinary $U(1)$ gauge theory
on commutative ${\R}^4$, it is easily shown that
non-singular instanton cannot exist:
For the ordinary 
$U(1)$ gauge field on commutative ${\R}^4$,
non-trivial instanton number is incompatible with the 
vanishing of field strength $F$ at infinity:
\bea
 -\frac{1}{8\pi^2} \int_{{\R}^4}FF
=- \frac{1}{8\pi^2} \int_{{\R}^4} d(AF)
=- \frac{1}{8\pi^2} 
\int_{S^3_{\mbox{\tiny infinity} }} (AF)
= 0. 
\ena
Therefore if there exists instanton,
there must be a singularity 
which gives a new surface term
other than $S^3$ at infinity.
Then
why $U(1)$ gauge field on noncommutative space
can have non-trivial
instanton charge ?
It becomes quite puzzling after the 
following considerations:
The difference between ordinary  
and noncommutative gauge theory
is the multiplication of field
(pointwise multiplication vs. 
star product) and
gauge field itself is written as 
smooth function on ${\R}^4$.
Actually, we can check that the explicit 
solution
is not singular.
On the other hand,
naively thinking,
the effect of noncommutativity
should be suppressed 
at long distance and does not give further 
contribution to the surface term at infinity.
So even in the 
noncommutative case
there seems no room for the non-singular
instanton. 
What's wrong with above arguments ?

The answer is: Above naive expectation is wrong.
The effect of noncommutativity 
does not vanish even at the long distance,
in the case of $U(1)$ gauge theory.
The reason is explained in the lectures.
\vs{5}

The organization of the lecture notes is as follows.
In section \ref{ncR} we review the calculational
techniques on noncommutative ${\R}^{2d}$.
The operator symbol reviewed here is
a fundamental tool 
throughout this note. 
Section \ref{sectopo} discusses
the first half of the main themes
in the lectures. After briefly reviewing the 
ADHM construction of instantons
on noncommutative ${\R}^4$,
we study explicit instanton solution and
clarify the topological origin of the 
integral instanton number in 
$U(1)$ gauge theory on noncommutative ${\R}^4$.
The relation between noncommutative
$U(1)$ instantons and projection operators is also reviewed,
where the projection operators act in Fock space
which is a representation space of noncommutative 
coordinates.
Section \ref{secIIB} discusses the latter half of 
the main themes in the lectures.
In the framework of IIB matrix model
we introduce a notion of gauge theory
restricted to the subspace
of the Fock space and 
consider the generalised gauge
equivalence relation.
Within this framework
we show that
the instanton number in 
$U(1)$ gauge theory is equal to the
dimension of the projection operator.
This gives an algebraic description for the
origin of the integral instanton number.
Finally we end with a summary and
speculations in section \ref{secsum}.
\newpage
\section{Calculus on Noncommutative %
${\R}^{2d}$}\label{ncR} 

\subsection{Noncommutative ${\R}^2$ }

First we consider
noncommutative ${\R}^2$,
since generalization to higher
dimension is straightforward.
Noncommutative ${\R}^2$ we shall consider
is described by coordinates
$\hat{x}^{\mu}$ ($\mu = 1,2$)
obeying the following commutation
relations:
\bea
\label{noncomx12}
[ \hat{x}^{1} ,  \hat{x}^{2}] 
= i \theta^{12}
= i \ell^{2},
\ena
where $\ell$ is a positive real 
constant and has dimension of length 
(the case $\theta^{12} < 0$ can be considered
in a similar manner).
We introduce
creation and annihilation 
operators by 
\bea
a^{\dagger } 
&\equiv& \frac{1}{\sqrt{2}\, \ell} 
\, \hat{z}, \qquad
\hat{z}
\equiv 
\hat{x}^2 + i \hat{x}^1 , \nn
a 
&\equiv& 
\frac{1}{\sqrt{2}\, \ell}
\, \hat{\bar{z}}, \qquad
\hat{\bar{z}} 
\equiv
\hat{x}^2 - i \hat{x}^1, \nn
& &\qquad
[a, a^{\dagger}] =1.
\ena
We realize the operators
$\hat{x}^{1} ,  \hat{x}^{2}$
in a Fock space ${\cal H}$
spanned by the basis $\left| n \right\ran$:
\bea
\left| n \right\ran 
\equiv \frac{1}{\sqrt{n!} }
(a^{\dagger })^n \left| 0\right\ran , \qquad
a \left| 0\right\ran = 0 .
\ena
The elements
of ${\cal H}$ are
called states or state vectors.
The dual space ${\cal H}^{\dagger}$
of ${\cal H}$ is spanned
by
$\left\lan n \right| $, \,
where 
$\left\lan n \right| \equiv
\left\lan 0 \right| a^{n}\frac{1}{\sqrt{n!}}, \,
\left\lan 0 \right| a^{\dagger } = 0$.
Let $\left|\,  \phi  \, \right\ran =
\sum_{n=0}^{\infty } \phi_n \left| n  \right\ran 
\in {\cal H}$,
where $\phi_n $ are complex numbers.
Then, the Hermite conjugate state is defined by
\bea
(\left| \phi \right\ran )^{\dagger}
\equiv   \sum_{n=0}^{\infty} \left\lan \, n \right| \phi_n^*
\quad .
\ena
The norm 
$ ||\,  \left| \phi \right\ran || \equiv
\left\lan \phi \! \right.\left| \phi \right\ran$ 
of the state $\left| \phi \right\ran \in {\cal H}$ 
is deduced from
\bea
\left\lan  0 \!\right. \left| 0 \right\ran 
=1.
\ena
The Hermite
conjugate operator
$\hat{O}^{\dagger }$ of $\hat{O}$
is defined by
$
\left\lan \phi \right| 
\hat{O}^{\dagger} 
=
(\hat{O}
\left| \phi \right\ran )^{\dagger}, \, 
\forall \left| \phi \right\ran \, 
\in {\cal H},\,
$.
The operator $\hat{O}$ 
is called Hermite
if it satisfies
$\hat{O}^{\dagger } = \hat{O}$.

The commutation relation (\ref{noncomx12})
has an automorphisms of the form
$ \hat{x}^{\mu} \mapsto \hat{x}^{\mu} + y^{\mu}$
(translation),
where $y^{\mu}$ is a commuting real number.
We denote the Lie algebra of this group
by ${\bf \underline{g}}$.
These automorphismes are
generated by unitary operator $U_y$:
\bea
U_y \equiv
\exp
[y^{\mu} \hat{\pa}_{\mu}] ,
\ena
where we have introduced
{\bf derivative operator}
$\hat{\pa}_{\mu}$ by
\bea
 \label{deri}
\hat{\pa}_{\mu} \equiv i B_{\mu\nu} \hat{x}^{\nu},
\ena
where $B_{\mu\nu}$ is a 
inverse matrix of $\theta^{\mu\nu}$. 
$\hat{\pa}_{\mu}$ satisfies following
commutation relations:
\bea
 \label{comdel}
[\hat{\pa}_{\mu}, \hat{x}^{\nu}] = \delta_{\mu}^{\nu}, \quad
[\hat{\pa}_{\mu}, \hat{\pa}_{\nu}]
= i B_{\mu\nu}.
\ena
One can check 
(cf. \ref{oprel2} in appendix \ref{oprel})
\bea
U_y\, \hat{x}^{\mu}\, U_y^{\dagger} =\,  \hat{x}^{\mu} + y^{\mu}.
\ena
We define derivative of operators by 
the action of \underline{\bf g}:
\bea
\label{del}
\pa_{\mu} \hat{O} \equiv
\lim_{\delta y^{\mu} \rightarrow 0}
\frac{1}{\delta y^{\mu} } 
\left(
U_{\delta y^{\mu}}
\hat{O} U_{\delta y^{\mu}}^{\dagger} - \hat{O}
\right)=
[\hat{\pa}_{\mu}, \hat{O} ].
\ena
Hence using (\ref{del}),
one can write the action of
derivative on
operator
algebraically.
The action of derivatives commutes:
\bea
\pa_\mu \pa_\nu \hat{O} - \pa_\nu \pa_\mu \hat{O}
=[\hat{\pa}_{\mu}, [\hat{\pa}_{\nu}, \hat{O} ]] 
- (\mu \leftrightarrow \nu)
= 0.
\ena


\subsection{
Operator Symbols}\label{opsymbol}

One can consider a one-to-one map
from operators to
ordinary c-number functions on ${\R}^2$
(operator symbols).
Under this map 
noncommutative operator multiplication is
mapped to star product.
The map from operators to
ordinary functions depends on
operator ordering prescription.
Weyl ordering is often used since it has
some convenient properties.
However 
the complex geometrical language is useful
in the description of instantons,
and in this case the normal ordering 
has some advantages.
Here we review Wick symbol
which corresponds to 
the normal ordering prescription.
As far as considering noncommutative ${\R}^{2d}$,
we can regard operators as fundamental objects, 
and regard operator symbols as mere representations.
Therefore we can choose any convenient 
operator ordering.
The important point is that
since operator symbols are ordinary functions
on ${\R}^2$,
we can discuss 
topological properties of noncommutative instantons
much in the same way as in the commutative case.

Let us consider normal ordered operator
of the form
\bea
 \label{Nop}
\hat{f}(\hat{x}) = 
\int \frac{d^2k}{(2\pi)^2} \, 
\tilde{f}(k) :  e^{ik\hat{x} } : \, ,
\ena
where $k\hat{x} \equiv k_{\mu}\hat{x}^{\mu}$.
$: \, \, :$ denotes the normal ordering.
For the 
operator valued 
function 
(\ref{Nop}),
the corresponding {\bf Wick symbol} 
is defined by
\bea
 \label{symb}
f_{\scriptscriptstyle{N}} (x) =
\int \frac{d^2k}{(2\pi)^2} \,
\tilde{f}(k) \ e^{ikx}\, , 
\ena
where $x^{\mu}$'s
are commuting coordinates of ${\R}^2$. 
We define $\Omega_N$ as a
map from operators to the Wick symbols:
\bea
\Omega_N(\hat{f}(\hat{x}) ) = 
f_{\scriptscriptstyle N}(x)
\equiv
\int \frac{d^2k}{(2\pi )^2}
\left( 2\pi \ell^2 
\mbox{Tr}_{\cal H} \left\{
\hat{f}(\hat{x})\ :e^{-ik\hat{x}}:
\right\} \right)
\,  e^{ikx} \quad .
\ena
Notice that from the relation
$\mbox{Tr}_{\cal H}
\bigl\{
: \exp \, ( ik\hat{x} ) : \bigr\}
= \frac{2\pi}{\ell^2} 
   \delta^{(2)}(k)$, %
it follows
\bea
 \label{intr}
2\pi \ell^2
\mbox{Tr}_{\cal H}\, \hat{f} (\hat{x})
=
\int d^2x\,  
f_{\scriptscriptstyle{N}} (x).
\ena  
The inverse map of $\Omega_N$ is given by
\bea
& &\Omega_N^{-1}(f(x)) = 
\hat{f}^{\scriptscriptstyle{N}}(\hat{x})
\equiv 
\int \frac{d^2k}{(2\pi)^2}
\left(
\int d^2x f(x)\  e^{-ikx}
\right)\,
: e^{ik\hat{x}}: \, .
\ena
The {\bf star product} 
of functions (corresponds
to the  normal ordering)
is defined by:
\bea
 \label{star}
f(x) \star_N g(x) \equiv
\Omega_N (\Omega_N^{-1} (f(x)) \Omega_N^{-1} (g(x))) \, .
\ena
Since 
\bea
 \label{starpro}
& &:e^{ik\hat{x}} :\, :e^{ik\hat{x}} :
=
e^{-\bar{w}\hat{z}} e^{w \hat{\bar{z}} }
e^{-\bar{w}' \hat{z} } e^{w'\hat{\bar{z}}}
=
e^{-\frac{\zeta}{2} w\bar{w}'}
e^{-(\bar{w} + \bar{w}') \hat{z} }
e^{(w + w' )\hat{\bar{z}} },
\ena
where
\bea
& &w  = \frac{i}{2}(k_2 + i k_1) ,
\ena
the explicit form of the star product is given by
\bea
f(z,\bar{z}) \star_N g(z,\bar{z})
=
\left.
e^{\frac{\zeta}{2} 
\frac{\pa }{\pa \bar{z} }
\frac{\pa}{\pa {z'} }
}
f(z,\bar{z}) 
g(z',\bar{z}')
\right|_{ z' = z, \bar{z}'=\bar{z} } \quad .
\ena
From the definition (\ref{star}), the star product
is associative:
\EQ
(f(x)\star_N g(x) \, ) \star_N h(x)
=
f(x)\star_N (g(x) \star_N h(x) \,).
\EN
If we use coherent states, the
expression of the
normal symbol becomes simpler.
The {\bf coherent state} 
$\left| \bar{z} \right\ran$
is an  eigen state
of annihilation operator 
$\hat{\bar{z}}$:
\bea
& &\hat{\bar{z}}\left| \bar{z} \right\ran
= \bar{z} \left| \bar{z} \right\ran .
\ena
As a notation we write the
dual state of $\left| \bar{z} \right\ran$ as
$\left\lan z \right|$ :
$\left\lan z \right| a^{\dagger}
= \left\lan z \right| z$.
Then the Wick symbol of operator
$\hat{f}$ is given by
\bea
 \label{fcohe}
f_{\scriptscriptstyle N} (z,\bar{z})
=
\left\lan z \right|
\hat{f}\left| \bar{z} \right\ran.
\ena
(\ref{fcohe}) follows from (\ref{Nop}), (\ref{symb})
and
\bea
\left\lan z \right|
: e^{ik\hat{x}}:
\left| \bar{z} \right\ran
&=&
\left\lan z \right|
e^{-\bar{w}\hat{z}} e^{w \hat{\bar{z}} }
\left| \bar{z} \right\ran
=
e^{-\bar{w}z} e^{w \bar{z} } \nn
&=&
e^{ikx} .
\ena
(we have normalised the coherent states as
$
\left\lan z \! \right.
\left| \bar{z} \right\ran = 1
$).
From (\ref{fcohe}) the Wick symbol 
$f_{\scriptscriptstyle N}(z,\bar{z})$
vanishes at
$z = w$ when the corresponding
operator $\hat{f}$ annihilates
$\left| \bar{w} \right\ran$
or
$\left\lan w \right|$:
\bea
 \label{van}
\hat{f}\left| \bar{w} \right\ran
= 0 \, \, \, \mbox{or} \, \,
\left\lan w \right| \hat{f}
= 0 \quad  \Longrightarrow 
\quad f_{\scriptscriptstyle N}(w,\bar{w})
 = 0 .
\ena

\subsubsection*{Examples}
First let us consider the
projection operator to the Fock vacuum
$\left|0 \right\ran$:
\bea 
 \label{Ex1}
\left|0 \right\ran\left\lan 0 \right|
= : e^{- a^{\dagger} a} : 
= : e^{- \frac{2}{\ell^2} \hz \hbz } : \quad .
\ena
Corresponding Wick symbol is
given by $e^{- \frac{1}{2\ell^2} z \bz }$.
Notice that this function 
concentrates  around 
the origin of ${\R}^2$ within
radius $\ell $.
It is also easy to obtain
\bea
 \label{Ex2}
\Omega_N 
( \left| m \right\ran\left\lan n \right| )
=
\frac{1}{\sqrt{m!}} 
\left(\frac{1}{\sqrt{2}\ell} z \right)^m 
e^{- \frac{2}{\ell} z \bz }
\left( \frac{1}{\sqrt{2}\ell}\bz \right)^n 
\frac{1}{\sqrt{n!}} \quad .
\ena
The Wick symbol
of the projection operator to the coherent state
$\left| \bw \right\ran$
is given by
\bea
 \label{Ex3}
\Omega_N(
\left| \bar{w} \right\ran\left\lan w \right|
)=
e^{- \frac{1}{2\ell^2} (z-w) (\bz-\bw) }\quad .
\ena
This function is obtained from
$\Omega_N (\left|0 \right\ran\left\lan 0 \right|)
=e^{- \frac{1}{2\ell^2} z \bz }$
by translation and
concentrates around $z=w$.

\subsection{Noncommutative ${\R}^{2d}$ }

Generalization to noncommutative ${\R}^{2d}$ may be
straightforward.
Coordinates on noncommutative ${\R}^{2d}$
obey the following commutation relations:
\bea
 \label{noncomxmn}
[\hat{x}^i,\hat{x}^j] = i \theta^{ij}.
\ena
Using
$SO(2d) $ rotations in ${\R}^{2d}$,
we can choose the coordinates
$\theta^{ij} = 0$ except 
$
\theta^{i (i+1)} = - \theta^{(i+1) i} \ne 0 , i = 1, \cdots d.
$
We assume
$\theta^{i (i+1)} > 0$ ($\theta^{i (i+1)} < 0$ case
can be treated similarly).
Then we define
creation and annihilation operators by
\bea
\hat{z}_i \equiv
\hat{x}_{i+1} + i \hat{x}_{i}, \quad
\hat{ \bar{z} }_i \equiv
\hat{x}_{i+1} - i \hat{x}_{i}. 
\ena
The noncommutative ${\R}^{2d}$
is described by 
operators acting in
the Fock space
${\cal H}$ spanned by the basis
$\left| n_1 , \cdots , n_d   \right\ran$,
where
\bea
& &\left| n_1 , \cdots , n_d   \right\ran \equiv
\Biggl\{ \Biggr.
\prod_{i=1 }^{d}
\frac{1}{\sqrt{n_i !}}
\left(\sqrt{\frac{1}{2\theta^{i(i+1)}}}\hat{z}_i
\right)^{n_i}
\Biggl. \Biggr\}
\left| 0,\cdots,0 \right\ran ,\nn
& &\hat{\bar{z} }_i \left| 0,\cdots,0 \right\ran =0 .
\ena
The correspondences
between 
the descriptions by
operators and by c-number functions are
given by
\bea
 \label{op-fn}
\hat{f} 
=
\int \frac{d^{2d}k }{ (2 \pi)^{2d} }
\, \tilde{f}(k) \, :\, e^{ik\hat{x}}\, : 
&\leftrightarrow  & 
f=
\int \frac{d^{2d}k }{ (2 \pi)^{2d} }
\, \tilde{f}(k)\, e^{ikx}, \nn
\hat{f}\hat{g}
&\leftrightarrow  & 
f \star_N g ,\nn
(2\pi)^d 
\sqrt{det \theta} \, 
\mbox{Tr}_{\cal H}
&\leftrightarrow  & 
\int d^{2d}x \quad .
\ena
Hereafter we will identify 
operator and corresponding Wick symbol
by eqs. (\ref{op-fn}): 
We use same letter for
operator and corresponding 
Wick symbol, as long as
it is not confusing.
For example we omit the hat $\hat{\,}$ from the
noncommutative coordinate $\hat{x}^{\mu}$
in the following.

\newpage
\section{Topological Charge of $U(1)$ %
Instantons on \\
Noncommutative ${\R}^4$}\label{sectopo}

\subsection{Gauge Theory on %
Noncommutative ${\bf R}^4$  }

Now let us consider noncommutative
${\R}^4$.
The coordinates 
$x^{\mu} \, \,  (\mu  = 1 ,\cdots , 4)$ of 
the noncommutative
${\R}^4$ obey
following
commutation relations:
\bea
 \label{noncomx}
[ x^{\mu} ,  x^{\nu}] = i \theta^{\mu\nu} , 
\ena
where $\theta^{\mu\nu}$ is real 
and constant.
We restrict ourselves
to the case
where $\theta^{\mu\nu}$ is {\em self-dual} and
set
\bea
 \label{theta}
\theta^{12} = \theta^{34} = \frac{\zeta}{4} 
\, \, ,
\ena
for simplicity.\footnote{%
As we will learn in section \ref{ncADHM}, the
condition needed for the
ADHM construction on noncommutative
${\R}^4$ is
$\theta^{12} + \theta^{34} 
=  \frac{\zeta}{2} $ \cite{SW}.
}
We introduce the complex 
coordinates by 
\bea
z_1 = x_2 + i x_1  , \quad z_2 = x_4 + i x_3 \,  .
\ena
Their commutation relations become
\bea
 \label{noncom}
 [z_1 , \bar{z}_1] 
=[z_2 , \bar{z}_2] 
= - \frac{\zeta}{2} \quad , \qquad 
\mbox{(others: zero)}.
\ena
We choose $\zeta >  0$.
We realize
$z$ and $\bar{z}$ 
as creation and annihilation operators
acting in a Fock space $\H$ spanned
by the basis
$\left| n_1 , n_2    \right\ran$:
\bea
\sqrt{\frac{2}{\zeta} } z_1 \left| n_1  , n_2    \right\ran
&=& \sqrt{n_1 +1} \left| n_1 + 1 , n_2    \right\ran , \quad
\sqrt{\frac{2}{\zeta} } \bar{z}_1 \left| n_1  , n_2    \right\ran
= \sqrt{n_1 } \left| n_1 -1 , n_2    \right\ran , \nn
\sqrt{\frac{2}{\zeta} } z_2 \left| n_1  , n_2    \right\ran
&=& \sqrt{n_2 +1} \left| n_1 , n_2 +1    \right\ran , \quad
\sqrt{\frac{2}{\zeta} }\bar{z}_2 \left| n_1  , n_2    \right\ran
= \sqrt{n_2} \left| n_1 , n_2 -1    \right\ran .
\ena
The action of the exterior derivative 
$d$ to the operator $O$ is 
defined as:
\bea
dO \equiv (\pa_{\mu} O )\,dx^{\mu}. 
\ena
Here $dx^{\mu}$'s
are defined in a usual way, i.e.
they commute with
$x^\mu$ and anti-commute among themselves:
$dx^{\mu}dx^{\nu} = -dx^{\nu}dx^{\mu}$.
The covariant derivative $D$ is
written as
\EQ
D = d+A.
\EN
Here $A= A_\mu dx^\mu$ 
is a $U(n)$ gauge field.\footnote{%
In this note the gauge field $A_\mu$ is
anti-Hermite.
}
The field strength of $A$ is given by
\bea
F \equiv D^2 = dA + A^2 
\equiv \frac{1}{2} F_{\mu\nu} dx^\mu dx^\nu.
\ena
We consider following 
Yang-Mills action:
\bea
 \label{SO}
S = \frac{1}{4g^2}
\left(
2\pi \frac{\zeta}{4}
\right)^2
\mbox{Tr}_{{\cal H}}\mbox{tr}_{U(n)}\,
F_{\mu\nu} F^{\mu\nu},
\ena
or we can rewrite it using Wick symbols:
\bea
 \label{Ss}
S = \frac{1}{4g^2}
\int
\mbox{tr}_{U(n)}\, F  * F,
\ena
where
$*$ is the Hodge star.\footnote{In this note
we only consider the case
where the metric 
on ${\R}^4$ is flat:
$g_{\mu\nu} = \delta_{\mu\nu}$.}
In (\ref{Ss}) multiplication of the
fields is understood  
as the star product.
The action (\ref{Ss})
is invariant under the following
$U(n)$ gauge transformation:
\bea
 \label{gauget}
A \rightarrow 
UdU^{\dagger} +
UAU^{\dagger}  .
\ena
Here $U$ is a unitary operator:
\bea
UU^{\dagger} = U^{\dagger}U =
\mbox{Id}_{\H} \otimes \mbox{Id}_{n} \quad .
\ena
Here $\mbox{Id}_{\H}$ is the identity operator
acting in $\H$ and $\mbox{Id}_{n}$ is 
the $n \times n$ identity matrix.
The gauge field $A$ is called
{\bf anti-self-dual}
if its field strength obeys the
following equation:
\bea
 F^+ \equiv
 \frac{1}{2} 
 (F + * F) = 0 .
\ena
Anti-self-dual gauge fields minimize
the Yang-Mills action (\ref{Ss}).
Instanton is an anti-self-dual
gauge field with finite
Yang-Mills action (\ref{Ss}).
The {\bf instanton number} is
defined by 
\bea
-\frac{1}{8\pi^2} \int \mbox{tr}_{U(n)}\,  F F \quad .
\ena
and takes integral value.
In ordinary 
$U(2)$ gauge theory,
integral instanton number
can be understood
topologically:
It is a winding number of gauge field
classified by $\pi_3(U(2))$.
However 
when the gauge group is $U(1)$, the
topological origin of the integral
instanton number
may seem unclear.
In the lectures
I will explain the origin
of the integral
instanton number 
in noncommutative $U(1)$ gauge theory. 

\subsection{ADHM Construction of Instantons
on Noncommutative ${\R}^4$ }\label{ncADHM}

ADHM construction is a way to obtain
instanton solutions on ${\R}^4$ from solutions of
some quadratic matrix equations \cite{ADHMconst}.
It was generalized to the case of
noncommutative ${\R}^4$ in \cite{NS}.
The steps in the ADHM construction
of instantons 
on noncommutative ${\R}^4$ 
in gauge group $U(n)$ with instanton number $k$
is as follows.

\begin{enumerate}

\item Matrices (entries are c-numbers):  
\bea
 B_1 , B_2 &:& k \times k \quad \mbox{complex matrices.} \nn
 I, J^{\dagger} &:& k \times n \quad \mbox{complex matrices.}
\ena

\item Solve the ADHM equations
\bea
\mu_{\R} &=&\zeta \,   \mbox{Id}_k 
\qquad \mbox{(real ADHM equation)},
\label{rADHM}\\ 
 \mu_{\C} &=& 0
\qquad \quad \mbox{(complex ADHM equation)} .
\label{cADHM}
\ena
Here $\mu_{\R}$ and $\mu_{\C}$
are defined by
\bea
 \label{ADHMzeta}
\mu_{\R}
 &\equiv& [B_1 , B_1^{\dagger}] 
+  [B_2 , B_2^{\dagger}] 
        + II^{\dagger} - J^{\dagger} J ,\\
\mu_{\bf C} &\equiv& [B_1 , B_2 ] + IJ  .
\ena  

\item Define $2k \times (2k + n)$ matrix
${\cal D}_{z}$ :
\bea
 \label{Dz}
& &{\cal D}_{z} \equiv
\left(
 \begin{array}{c}
   \tau_{z} \\
   \sigma_{z}^{\dagger }
 \end{array}
\right) , \nn
& & \tau_{z} \equiv
(\, B_2 - {z}_2 ,\, B_1 - {z}_1 , \, I \, ), \nn
& & \sigma_{z}^{\dagger}  \equiv
( \, - (B_1^{\dagger}-\bar{{z}}_1) , \,
  B_2^{\dagger} - \bar{{z}}_2  , \, J^{\dagger} \, ) .
\ena
Here $z$'s are noncommutative
{\em operators}.

\item Look for all the solutions to the equation 
\bea
 \label{zeroPsi}
{\cal D}_{z} \Psi^{(a)} = 0  \quad
( a = 1, \ldots , n),
\ena
where $\Psi^{(a)} $ is a $2k+n$ dimensional vector
and its entries 
are {\em operators}: \ %
Here we impose following normalization condition
for $\Psi^{(a)} $:
\bea
 \label{norm}
\Psi^{(a)\dagger}\Psi^{(b)} = \delta^{ab} \mbox{Id}_{\H} \quad .
\ena
In the following we will call these zero-eigenvalue vectors 
$\Psi^{(a)}$  {\bf zero-modes}.

\item Construct gauge field by the formula
\bea
\label{ncA} 
A^{ab}_\mu
= \Psi^{(a) \dagger } \pa_\mu \Psi^{(b)} , 
\ena
where $a$ and $b$ become indices
of $U(n)$ gauge group.
Then this gauge field is anti-self-dual.
\end{enumerate}
In the proof of the anti-self-duality,
following equations are important.
\bea
\label{key}
\tau_{z} \tau_{z}^{\dagger}
&=&  \sigma_{z}^{\dagger} \sigma_{z} := \Box_{z}, 
\label{k1}\\
\tau_{z} \sigma_{z} &=& 0 , \label{k2}\\
\Psi^{(a)\dagger}\Psi^{(b)} &=& \delta^{ab} \mbox{Id}_{\H}
\label{k3}.
\ena
The equation (\ref{k1}) is equivalent to the real
ADHM equation provided that the coordinates
are noncommutative as in $\ref{noncom}$,
or more generally
$[z_1,\bz_1] + [z_2,\bz_2] = -\zeta$.
This is what Nekrasov and Schwarz found in their
pioneering work \cite{NS}.\footnote{%
The relation between
parameter $\zeta$ in eq.(\ref{rADHM}) 
and NS-NS B-field background
in string theory
was first pointed out in \cite{ABS}.
}
The equation (\ref{k2}) is equivalent to the 
complex ADHM equation.
The equation (\ref{k3}) is simple
but necessary (however we can relax
this condition by extending the notion
of gauge equivalence. See section \ref{secIIB}). 
Let us check that the field strength
constructed from 
(\ref{ncA}) is really anti-self-dual:
\bea
F&=&
dA + A^2 \nn
&=&
d (\Psi^{\dagger} d \Psi )
+ (\Psi^{\dagger} d \Psi)( \Psi^{\dagger} d \Psi) 
\label{adsf1}\nn
&=& 
d \Psi^{\dagger} (1- \Psi\Psi^{\dagger})d \Psi.
\label{adsf2}
\ena
In the above we have suppressed 
$U(n)$ indices.
Here we have already used the condition (\ref{k3}).
One of the key ingredients
in the ADHM construction
is that
$(1- \Psi\Psi^{\dagger})$ is a projection
acting on ${\C}^{2k + n} \otimes \H$ 
and project out 
the space of zero-modes,
since $\Psi\Psi^{\dagger}$ is
a projection to the space of zero-modes
(because of the normalization condition (\ref{k3})).
Hence it can be rewritten as
\bea
 \label{projADHM}
1- \Psi\Psi^{\dagger} &=& 
{\cal D}_z^{\dagger}
\frac{1}{ {\cal D}_z {\cal D}_z^{\dagger}  }
{\cal D}_z  \quad .
\ena
Indeed 
$ {\cal D}_z^{\dagger}
\frac{1}{ {\cal D}_z {\cal D}_z^{\dagger}  }
{\cal D}_z$ is a projection 
that project out the zero-modes of ${\cal D}_z$:
\bea
\left( {\cal D}_z^{\dagger}
\frac{1}{ {\cal D}_z {\cal D}_z^{\dagger}  }
{\cal D}_z \right)^2
=
 {\cal D}_z^{\dagger}
\frac{1}{ {\cal D}_z {\cal D}_z^{\dagger}  }
{\cal D}_z
 {\cal D}_z^{\dagger}
\frac{1}{ {\cal D}_z {\cal D}_z^{\dagger}  }
{\cal D}_z
=
 {\cal D}_z^{\dagger}
\frac{1}{ {\cal D}_z {\cal D}_z^{\dagger}  }
{\cal D}_z \quad .
\label{DDDD}
\ena
(\ref{DDDD}) can be written as
\bea
{\cal D}_z^{\dagger}
\frac{1}{ {\cal D}_z {\cal D}_z^{\dagger}  }
{\cal D}_z
&=&
\tau^{\dagger}_z \frac{1}{\, \tau_z\tau^{\dagger}_z} \tau_z
+
\sigma_z \frac{1}{\, \sigma_z^{\dagger} \sigma_z}
\sigma_z^{\dagger } \nn
&=&
\tau^{\dagger}_z \frac{1}{\, \, \Box_z} \tau_z
+
\sigma_z \frac{1}{\, \, \Box_z}
\sigma_z^{\dagger },
\ena
where we have used the notations in (\ref{key}).
Since $\tau_z \Psi = \sigma_z^{\dagger } \Psi =0$
by definition (\ref{zeroPsi}), it follows that
$\tau_z  d \Psi = -d \tau_z \Psi,\, 
 \sigma_z^{\dagger} d \Psi 
= - d \sigma_z^{\dagger} \Psi$.
Hence
\bea
\label{cFS}
F  
&=&
d \Psi^{\dagger} (1- \Psi\Psi^{\dagger})d \Psi \nn
&=&
d \Psi^{\dagger}
\left( \tau^{\dagger}_z \frac{1}{\, \, \Box_z} \tau_z
+
\sigma_z \frac{1}{\, \, \Box_z}
\sigma_z^{\dagger } \right)
d \Psi \nn
&=&
 \Psi^{\dagger}
\left( d\tau^{\dagger}_z \frac{1}{\, \, \Box_z} d\tau_z
+
d\sigma_z \frac{1}{\, \, \Box_z}
d\sigma_z^{\dagger } \right)
\Psi \nn
&=&
\begin{array}{ccc}
\bigl(\, \psi_1^{\dagger} \,& \,
 \psi_2^{\dagger} \,& \, \xi^{\dagger} \, 
\bigr) \\
   &  &  \\
   &  &  
\end{array} \! \!
\left(
 \begin{array}{ccc}
dz_1 \frac{1}{\, \, \Box_z}d\bar{z}_1 
+ d\bar{z}_2 \frac{1}{\, \, \Box_z} dz_2   & 
  -dz_1 \frac{1}{\, \, \Box_z} d\bar{z}_2 
       + d\bar{z}_2\frac{1}{\, \, \Box_z} 
dz_1 & 0 \\
-dz_2 \frac{1}{\, \, \Box_z} d\bar{z}_1 
+ d\bar{z}_1\frac{1}{\, \, \Box_z} dz_2 & 
  dz_2 \frac{1}{\, \, \Box_z} d\bar{z}_2 
  + d\bar{z}_1 \frac{1}{\, \, \Box_z} dz_1 & 0 \\
 0 & 0 & 0
 \end{array}
\right) 
\left(
\begin{array}{c}
\psi_1 \\
\psi_2 \\
\xi
\end{array}
\right) \nn
&\equiv& F^{-}_{\mbox{\tiny ADHM}} \quad .
\ena
where we have written
\bea
& &\Psi \equiv
\left(
\begin{array}{c}
\psi_1 \\
\psi_2 \\
\xi
\end{array}
\right) \equiv
\left(
\begin{array}{ccc}
 & & \\
\Psi^{(1)} & \cdots & \Psi^{(n)} \\
 & &
\end{array}
\right) , \qquad
\begin{array}{c}
\psi_1 : k \times n \, \, \mbox{matrix.}\\
\psi_2 : k \times n \, \, \mbox{matrix.}\\
\, \,  \xi \, \,  : n \times n\, \, \mbox{matrix.}
\end{array}
\ena
$F^{-}_{\mbox{\tiny ADHM}}$ is 
anti-self-dual:
$F_{z_1\bz_1}+ F_{z_2\bz_2} = 0$,
$F_{z_1 z_2} = 0$.


Now let us consider the large 
$r \equiv \sqrt{x^\mu x_\mu}$ 
behaviour
of the gauge field thus constructed.
From the form of ${\cal D}_z$ in (\ref{Dz}),
the asymptotic form of the zero-mode $\Psi$ becomes
\bea
 \label{asPsi}
 \Psi =
 \left(
 \begin{array}{c}
 0 \\
 0 \\
 \xi
 \end{array}\right)
+ O(r^{-1}) \quad (r \rightarrow \infty) .
\ena
Then the asymptotic
form of the gauge field becomes
\bea
 \label{asA}
A_\mu =
\xi^{\dagger} \pa_\mu \xi +
O(r^{-2}) \quad (r \rightarrow \infty).
\ena
From (\ref{asA}) we can observe that only
the third factor $\xi$ of $\Psi$
is relevant for the asymptotic
behaviour of the gauge field.


\subsection{$U(1)$ One-Instanton Solution and %
Its Topological Charge}\label{oneone}

Now that we have reviewed the
ADHM construction of instantons
on noncommutative ${\R}^4$,
we can construct explicit instanton solutions.
The simplest solution is 
$U(1)$ one-instanton solution.
In this case $B_1$ and $B_2$ are $1\times1$ matrices,
i.e. complex numbers. Therefore commutators with 
$B_1$ and $B_2$ are automatically give zero and
the solution to the ADHM equation
(\ref{ADHMzeta})
is given by
\bea
 \label{1-1}
\quad I = \sqrt{\zeta}\, ,\quad \, J = 0,
\ena
with $B_1$ and $B_2$ arbitrary.
$B_1$ and $B_2$ are parameters
that represent the position of instanton.
Due to the translational invariance 
on noncommutative ${\R}^4$,
it is enough if we consider $B_1 = B_2 = 0$ solution.
There is a solution $\tilde{\Psi}_0 $ to 
the equation 
${\cal D}_z \tilde{\Psi}_0 = 0 $ :
\bea
 \label{1-1zeroPsi}
\tilde{\Psi}_0 =
\left(
 \begin{array}{c}
  \tilde{\psi_1} \\
  \tilde{\psi_2} \\
  \tilde{\xi}
 \end{array}
\right) = 
\left(
 \begin{array}{c}
\sqrt{\zeta} \bar{z}_2 \\
\sqrt{\zeta} \bar{z}_1 \\ 
\left( z_1\bar{z}_1 + z_2 \bar{z}_2 \right) 
 \end{array}
\right)  .
\ena
Notice that all the components of
$\tilde{\Psi}_0 $ annihilate 
$\left| 0,0 \right\ran$.
As a consequence 
$\tilde{\Psi}_0^{\dagger} \tilde{\Psi}_0 
= (z_1\bar{z}_1 + z_2\bar{z}_2 )
(z_1\bar{z}_1 + z_2\bar{z}_2 + \zeta)$
annihilates $\left| 0,0 \right\ran$.
This means 
$\tilde{\Psi}_0^{\dagger} \tilde{\Psi}_0$
does not have its inverse operator and
we cannot normalize the zero-mode
$\tilde{\Psi}_0$.
This is a problem since
we cannot follow the 
steps in ADHM construction,
see eq.(\ref{norm}) and eq.(\ref{k3}). 
What should we do ?

The solution of this problem can be found
if we notice that
the dimension of the Fock space $\H$
is infinite.
Before explaining the reason in the case
of instanton,
let us recall the famous story
which shows the
mysterious nature of infinity.

\begin{figure}
\begin{center}
 \leavevmode
 \epsfxsize=130mm
 \epsfbox{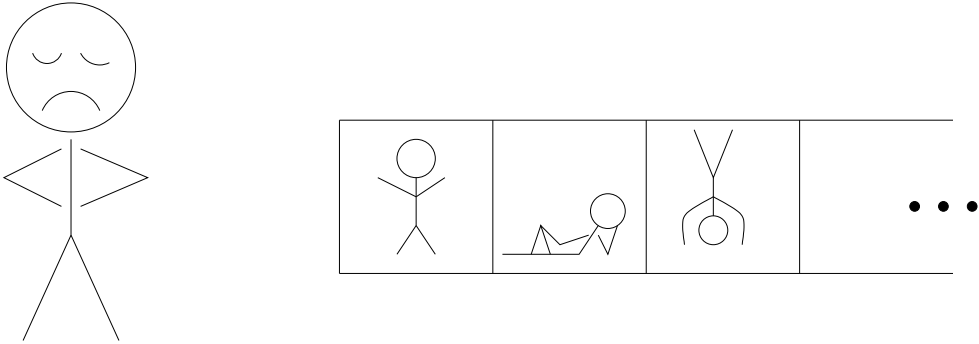}
\end{center}
\caption{Can you make another room for the 
traveller ?}
\label{noroom} 
\end{figure}

\begin{figure}
\begin{center}
 \leavevmode
 \epsfxsize=130mm
 \epsfbox{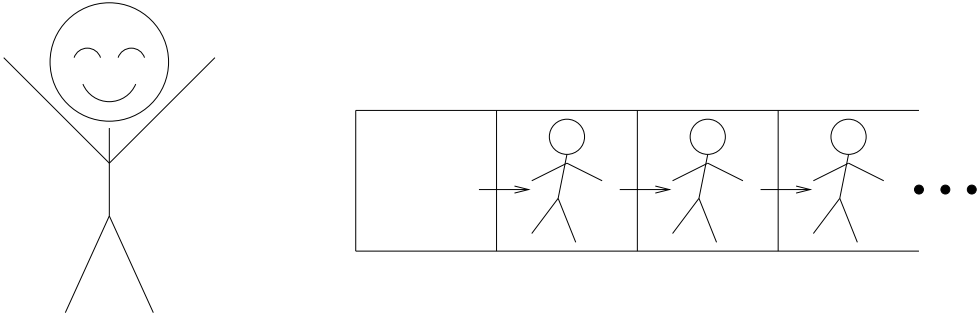}\\
\end{center}
\caption{We have a room for the travellar and now
he is happy !}
\label{oneroom} 
\end{figure}

\begin{figure}
\begin{center}
 \leavevmode
 \epsfxsize=130mm
 \epsfbox{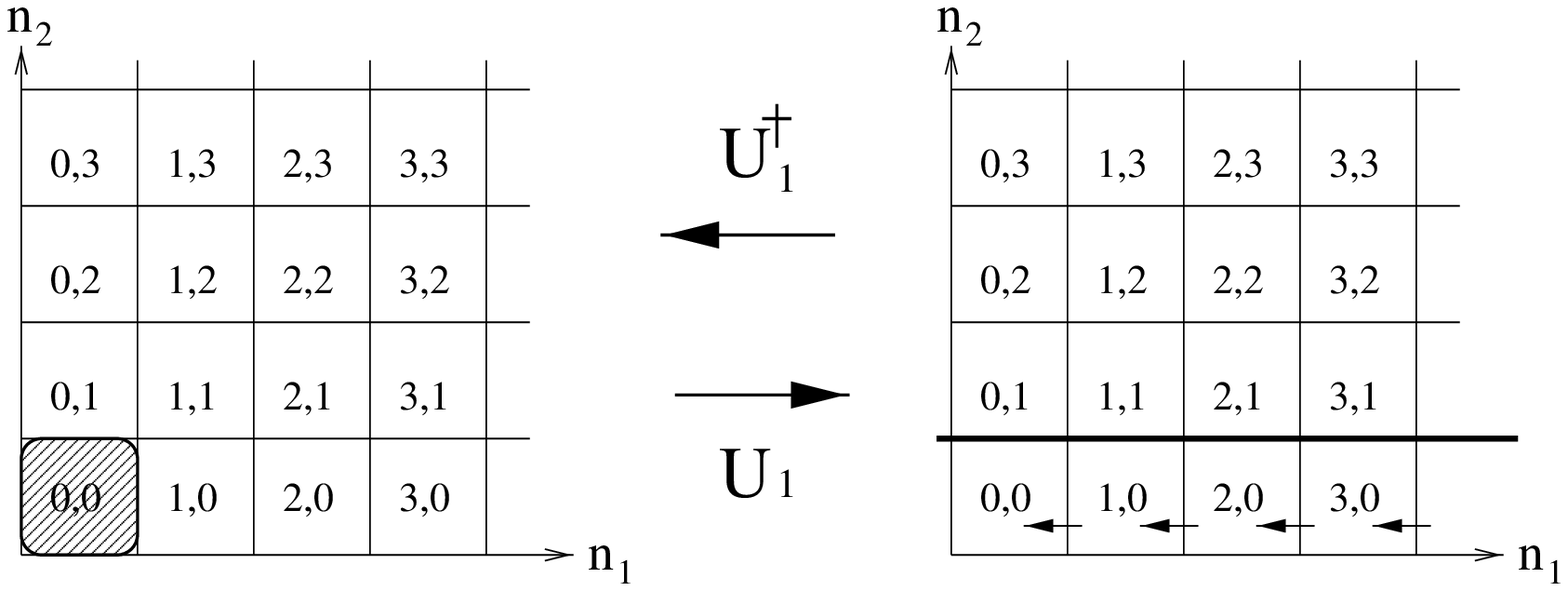}\\
\begin{picture}(0,0)
\put(-130,-10){\Large $p\H$}
\put(90,-10){\Large $\H$}
\end{picture}
\end{center}
\caption{
}
\label{figaroMvN} 
\end{figure}

There was a hotel in some far place, and
there was an infinite number of rooms
in that hotel.
One day all rooms were filled.
Then a traveller who wanted
to stay at that hotel came, see fig.\ref{noroom}. 
Can you make another room for the traveller ?
The answer may be apparent from fig.\ref{oneroom}:
We require every guest
to move to the next room.
%
%
Tracing back this argument,
we can find a solution of the problem 
in the case of instanton.
We want to eliminate the zero-eigenvalue
component of 
$\tilde{\Psi}_0^{\dagger} \tilde{\Psi}_0$,
that is,
$\left| 0,0 \right\ran
\left\lan 0,0\right|$.
This corresponds to the 
empty room in fig. \ref{oneroom}.
What we should do is to find a
operator $U$ which 
gives one-to-one map
from $p\H$ to $\H$, where
$p = 1 - 
\left| 0,0 \right\ran
\left\lan 0,0\right| $ .
Such operator $U$
satisfies \cite{Ho}\cite{mine2}
\begin{eqnarray}
 \label{MvNU}
 UU^{\dagger} = 1, \quad 
 U^{\dagger}U = p =
1 - 
\left| 0,0 \right\ran
\left\lan 0,0\right|   .
\end{eqnarray}
One example of such operator $U_1$
is given by
\bea
 \label{U1}
 U_1 &=& 
 ( 1- \left| 0 \right\ran
 \left\lan 0\right|_2)
 + 
\sum_{n_1 =0}^{\infty} \left| n_1,0 \right\ran
 \left\lan n_1+1 ,0\right| \nn
&=&
  (1- \left| 0 \right\ran
 \left\lan 0\right|_2)
 + \left| 0 \right\ran
 \left\lan 0\right|_2 
\frac{1}{\sqrt{\hn_1 +1} }  a_1, \nn
\ena
where we have defined
\bea
\left| m \right\ran
 \left\lan n \right|_2
\equiv
\sum_{n_1 = 0}^{\infty}
 \left| n_1, m \right\ran
 \left\lan n_1, n \right| ,
\ena
see fig.\ref{figaroMvN} .
%
%
%
%
To construct a zero-mode
which is normalized as in
(\ref{norm}),
we first normalize
the zero-mode $\tilde{\Psi}_0$ in the subspace of the
Fock space where $\left| 0,0 \right\ran$ 
is projected out:
\begin{eqnarray}
 \label{FPsi0}
 \Psi_0 \equiv \tilde{\Psi}_0 
(\tilde{\Psi}_0^{\dagger} \tilde{\Psi}_0)^{-1/2}
\equiv
\left(
 \begin{array}{c}
  (\psi_1)_0 \\
  (\psi_2)_0 \\
  \xi_0
 \end{array}
\right).
\end{eqnarray}
Here $(\tilde{\Psi}_0^{\dagger} \tilde{\Psi}_0)^{-1/2}$
is defined in $p \H$:
\bea
 \label{invtPsi}
(\tilde{\Psi}_0^{\dagger} \tilde{\Psi}_0)^{-1/2}
\equiv
\frac{2}{\zeta}
\sum_{(n_1,n_2)\ne(0,0)}
\frac{1}{(n_1+n_2)(n_1+n_2+2)}
\left|n_1,n_2\right\ran
\left\lan n_1,n_2 \right|.
\ena
Notice that
(\ref{invtPsi}) is well defined
since it is defined in $p \H$
where 
$\left|0,0\right\ran
\left\lan 0,0 \right|$
is projected out.
$\Psi_0$ is normalized as 
\begin{eqnarray}
 \Psi_0^{\dagger} \Psi_0   = p 
= 1 - \left| 0,0 \right\ran
 \left\lan 0,0\right| .
\end{eqnarray}
Then we define $\Psi$ using $U_1$:
\begin{eqnarray}
 \label{PoU}
 \Psi = \Psi_0 U_1^{\dagger}\quad .
\end{eqnarray}
$\Psi$ is normalized as in (\ref{norm}):
\bea
\Psi^{\dagger} \Psi =
U_1 \Psi_0^{\dagger} \Psi_0 U_1^{\dagger}
=
U_1 (1 - \left| 0,0 \right\ran\left\lan 0,0\right|)
U_1^{\dagger}
=
1.
\ena
Hence the gauge field
\bea
 \label{Ainst}
A_\mu = \Psi^{\dagger} \pa_\mu \Psi ,
\ena
is anti-self-dual.

\begin{figure}
\begin{center}
 \leavevmode
 \epsfxsize=60mm
 \epsfbox{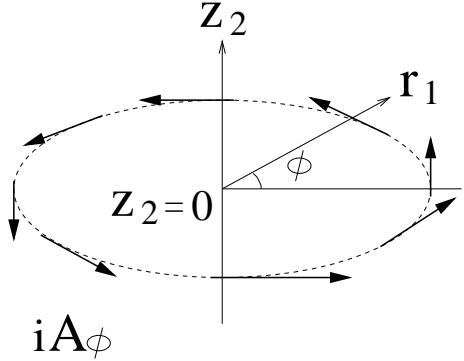}\\
\end{center}
\caption{At large $r_1$ the gauge field
concentrates around $z_2 = 0$ plane.
After the integration over $z_2$
the instanton number reduces to the 
winding number of gauge field around $S^1$ on 
$z_1$ plane.}
\label{wind} 
\end{figure}

Now let us calculate the
instanton number.
It is written as a
surface term in
the same way as in the commutative case: 
\bea
-\frac{1}{8 \pi^2}
\int_{{\R}^4} 
F F
=
-\frac{1}{8 \pi^2}
\int_{{\R}^4} 
d K 
=
-\frac{1}{8 \pi^2}
\int_{\mbox{surface at }\infty} K.
\ena
Here
\bea
 K \equiv AdA + \frac{2}{3} A^3
   = A F - \frac{1}{3} A^3 .
\ena
Now let us consider the asymptotic behaviour
of the gauge field.
Since the third factor $\xi_0$ of
$\Psi_0$ goes to $1$ for 
$r\rightarrow \infty$,
the asymptotic form of 
the gauge field is determined 
by the asymptotic form of $U_1$ (see (\ref{asA})):
\begin{eqnarray}
A_\mu \rightarrow
U_1 \pa_\mu U^{\dagger}_1 \quad(r \rightarrow \infty).
\end{eqnarray}
Notice that the Wick symbol of
$1 - p_2$ appearing in $U_1$
does not vanish
around the $z_2 = 0$ plane even if we 
take $r$ to infinity:
\begin{eqnarray}
 \label{Wick00}
 \Omega_N (1 - p_2) = 
  e^{- \frac{2}{\zeta} z_2 \bar{z}_2 }.
\end{eqnarray}
From (\ref{Wick00}) we observe that
the gauge field exponentially damps
as $r_2 \equiv \sqrt{(x^3)^2 + (x^4)^2} \rightarrow \infty$.
Therefore in order to calculate the surface term we can choose
$r_1 = R_1 \, \,(\mbox{const.})$ surface
and take $R_1$ to $\infty$ .
\bea
\int_{{\R}^4}  d K
=
\int_{r_1 = R_1} K
=
-\frac{1}{3}
\int_{r_1 = R_1} A^3 \quad (R_1 \rightarrow \infty).
\ena
Let us consider 
$r_1 \rightarrow \infty$ behaviour
of $U_1$. 
We introduce 
the polar coordinates on $z_1$ plane:
\EQ
z_1 = r_1 e^{i\phi}.
\EN
Then in the $r_1 \rightarrow \infty$ limit
the Wick symbol of 
$\frac{1}{\sqrt{\hn_1 +1} }  a_1 $
becomes
\bea
 \label{asna}
 \Omega_N \left(\frac{1}{\sqrt{\hn_1 +1} }  a_1 \right)
  \rightarrow e^{-i\phi} \quad (r_1 \rightarrow \infty).
\ena
eq. (\ref{asna}) essentialy
gives the
topological origin of the
instanton number.
From (\ref{asna}) we obtain
\bea
U_1 \rightarrow 
p_2 + (1-p_2) e^{-i\phi} =
(1-p_2) (e^{-i\phi} -1)- 1
\quad
(r_1 \rightarrow \infty).
\ena
The large
$r_1$ behaviour of the gauge field becomes
\begin{eqnarray}
 A_{r_1}
&\rightarrow& 0 ,\\
 A_\phi 
&\rightarrow& U_1 \pa_\phi U_1^{\dagger} 
     = i (1-p_2) 
= i  \left| 0 \right\ran \left\lan 0 \right|_2 ,\\
 A_{z_2}
&\rightarrow& U_1 \pa_{z_2} U_1^{\dagger} 
= - e^{i\phi} (e^{-i\phi} -1) 
\sqrt{\frac{2}{\zeta}}
  \left| 0 \right\ran \left\lan 1 \right|_2 ,\\
  A_{\bar{z}_2}
&\rightarrow& U_1 \pa_{\bz_2} U_1^{\dagger} 
    = e^{-i\phi}  (e^{i\phi} -1)
     \sqrt{\frac{2}{\zeta}}
       \left| 1 \right\ran \left\lan 0 \right|_2 
\qquad (r_1 \rightarrow \infty).
\end{eqnarray}
It is convenient to 
use Wick symbol for the calculation in $z_1$ plane
and operator calculus for
$z_2$ plane:
\bea
\int_{r_1 = R_1} A^3 
&=&
\int_0^{2\pi} d\phi \, 
2i \left( 2\pi\frac{\zeta}{4}  \right)\mbox{tr}_2
 A_\phi A_{z_2} A_{\bar{z}_2} \times 3 
\quad (R_1 \rightarrow \infty).
\ena
Notice that 
$\int dz_2 d\bz_2 
= \int 2i dx_1 dx_2
= 2i \left( 2\pi\frac{\zeta}{4}  \right)\mbox{tr}_2$,
and 
$\mbox{tr}_2 
A_\phi A_{\bar{z}_2}  A_{z_2}= 0$
(give care to the ordering).
Thus
\bea
\int_{r_1 = R_1 \rightarrow \infty} A^3 &=& 
2 \pi \int_0^{2\pi} d\phi \,
(e^{i\phi} -1) (e^{-i\phi} -1) \times 3 \nn
&=& 24 \pi^2 ,
\ena
and the instanton number becomes
\bea
-\frac{1}{8 \pi^2}
\int F F
=
\frac{1}{8 \pi^2} \int \frac{1}{3} A^3
= 1 .
\ena
The important point is that
since the gauge field only
has 
$\left| 0 \right\ran \left\lan 0 \right|_2 $,
$\left| 1 \right\ran \left\lan 0 \right|_2 $ and
$\left| 0 \right\ran \left\lan 1 \right|_2 $
components,
the trace over $n_2$
essentially reduces the 
the instanton number to the 
winding number of 
the gauge field $A_\phi$  
around $S^1$  on $z_1$ plane (fig.\ref{wind}).
Thus the instanton number is 
characterized by $\pi_1(U(1))$.
Indeed in $\phi$ integration,
only the constant part contributes,
and $A_{z_2}$ and $A_{\bz_2}$ only 
give appropriate numerical factor.
This explains the origin of the
integral instanton number
for the case of noncommutative
$U(1)$ instanton.

Of course this explanation is gauge dependent:
For example, we could have used 
\bea
 \label{U2}
 U_2 &\equiv& 
  1  -\left|0 \right\ran
 \left\lan 0 \right|_1
 +  \sum_{n_2} \left| 0, n_2 \right\ran
 \left\lan 0, n_2 +1 \right| \nn
&=&
  1  -\left| 0 \right\ran
 \left\lan 0 \right|_1
 + 
\left| 0 \right\ran
 \left\lan 0 \right|_1
\frac{1}{\sqrt{\hn_2 +1} }  a_2 ,
\ena
where we have defined
\bea
 \left|m  \right\ran
 \left\lan n \right|_1 
\equiv \sum_{n_2 = 0}^{\infty}
 \left|m , n_2 \right\ran
 \left\lan n , n_2\right| ,
\ena
instead of $U_1$. 
$U_2$ gives a winding number 
around
$S^1$ in $z_2$ plane.
The important point is that
although the choice of $S^1$
depends on gauge,
any operator that satisfies
(\ref{MvNU})
necessarily introduces 
winding number around $S^1$.


\subsection{%
Noncommutative $U(1)$ Instantons and
Projection Operators}\label{tideal}

In the previous subsection
we have seen that
the noncommutative $U(1)$ instanton 
at origin 
has close relation with
the projection operator
$\left| 0,0 \right\ran \left\lan 0,0\right|$.
In this subsection 
we will investigate this
relation between 
noncommutative $U(1)$ instantons
and projections
for the general multi-instanton solutions.
For this purpose we first 
study the solution of the 
equation (\ref{zeroPsi}).
Let us consider the solution to the equation
\bea
 \label{zeroV}
{\cal D}_z \left| {\cal U} \right\ran = 0 , 
\ena
where 
$ \left| {\cal U} \right\ran 
\in ({\cal H}\otimes {\C}^k ) \oplus
    ({\cal H} \otimes {\C}^k) \oplus  
    {\cal H}$, 
i.e.
the components of 
$ \left| {\cal U} \right\ran $
are {\it vectors} 
in the Fock space ${\cal H} $.
We call $ \left| {\cal U} \right\ran $
{\bf vector zero-mode} and
call $\Psi $ 
in (\ref{zeroPsi}) 
{\bf operator zero-mode}.
We can construct
operator zero-mode
if we know all the vector zero-modes.
The advantage of considering
vector zero-modes
is that we can relate
them to
the ideal\footnote{%
Definition of ideal
is given in appendix \ref{appideal}. }
discussed in 
\cite{Nakaj}\cite{LecNakaj}.
The point is that we can
regard vector zero-modes
as holomorphic vector bundle 
described
in purely commutative terms.
This is because
we can identify states in Fock space
and polynomials on ${\bf C}^2$:
\bea
 \label{identify}
f(z_1,z_2) \left| 0,0 \right\ran
\longleftrightarrow f(z_1,z_2) \cdot 1 .
\ena
The $z_1$ and $z_2$ in the
left hand side are operators,
but as long as we are considering only 
creation operators, 
we can identify them with c-number-coordinates
because they commute with each other.

Let us write
\bea
\label{veczero}
\left| {\cal U} \right\ran = 
\left(
 \begin{array}{c}
 \left| u_1 \right\ran \\
  \left| u_2 \right\ran \\
  \left|\, f \, \right\ran 
\end{array}
\right) ,
\qquad
 \begin{array}{l}
 \left| u_1 \right\ran \equiv
   u_1(z_1,z_2) \left|\,  0,0 \,\right\ran ,\\
\left| u_2 \right\ran \equiv
  u_2(z_1,z_2)\left|\, 0,0\, \right\ran ,\\
  \left|\, f \, \right\ran 
\equiv  f (z_1,z_2)\left|\, 0,0\, \right\ran . 
\end{array}
\ena
where
$
\left| u_1 \right\ran ,
\left| u_2 \right\ran  \in 
 {\cal H}\otimes {{\C}^k}$
i.e. they are
vectors in ${\bf C}^k$ 
and state vectors in ${\cal H}$,
and
$
\left|\, f \, \right\ran \in {\cal H}
$.
The space of the solutions 
of (\ref{zeroV}):
$
\ker {\cal D}_z =
\ker \tau_z \cap
 \ker \sigma_z^{\dagger } 
\simeq 
\ker \tau_z/ \mbox{Im}\, \sigma_z $\footnote{%
Notice that 
since
$ \tau_z \sigma_z = 0$ ((\ref{key})),
$\ker \tau_z/ \mbox{Im}\, \sigma_z $ is 
well defined. 
$\ker \tau_z \cap
 \ker \sigma_z^{\dagger } 
\simeq 
\ker \tau_z/ \mbox{Im}\, \sigma_z $ is understood
as follows: The condition
$\sigma_z^{\dagger }
\left|\, {\cal U} \, \right\ran  = 0$ fixes
the ``gauge freedom" 
mod Im $\sigma_z$
in $\ker \tau_z/ \mbox{Im}\, \sigma_z $. 
See appendix \ref{appideal}. 
}
is isomorphic to the ideal ${\cal I}$ 
defined by 
\bea
 \label{ideal}
{\cal I} 
= \Bigl\{\, f(z_1,z_2) \, 
\Bigm|\, \, f(B_1, B_2) = 0 \, \, \Bigr\} ,
\ena
where $B_1$ and $B_2$ together with
$I$ and $J$ give a solution to (\ref{ADHMzeta}).
In $U(1)$ case, one can show 
$J = 0$, and
the isomorphism is given by the inclusion 
of the third factor in (\ref{veczero}) 
\cite{Nakaj}\cite{LecNakaj} 
(See appendix \ref{appideal}).
\bea
\label{inclusion}
\ker \tau_z/ \mbox{Im}\, \sigma_z\,
 \hookrightarrow {\cal O}_{{\bf C}^2}\,:\quad
\left| {\cal U} \right\ran = 
\left(
 \begin{array}{c}
 \left| u_1 \right\ran \\
  \left| u_2 \right\ran \\
  \left|\, f \, \right\ran 
\end{array}
\right)\,  \hookrightarrow f(z_1,z_2),
\ena
where ${\cal O}_{{\bf C}^2}$ denotes
ring of polynomials on ${\bf C}^2$.
We define {\bf ideal state} by 
the identification (\ref{identify})
\bea
\left| \varphi \right\ran 
\in \mbox{ideal states of ${\cal I}$ }
\Longleftrightarrow
\exists f( z_1, z_2 ) \in {\cal I} , \quad
\left| \varphi \right\ran =
f( z_1, z_2 ) \left| 0,0 \right\ran ,
\ena
and denote the 
space of all the ideal states by
${\cal H}_{{\cal I} }$.
We define 
projection operator $p_{\cal I}$
as a projection to the
space of ideal ${\cal H}_{{\cal I} }$:
\bea
p_{\cal I} {\cal H} = {\cal H}_{{\cal I} }.
\ena
We define ${\cal H}_{/{\cal I} }$
\footnote{The meaning of this notation
is as follows:
${\cal H}_{/{\cal I} }$ corresponds to
${\cal O}_{{\C}^2}/ {\cal I }$, where
${\cal O}_{{\C}^2} $ is the ring of polynomials
on ${\C}^2$.}
as a 
subspace of ${\cal H}$
orthogonal to ${\cal H}_{{\cal I} }$ :
\bea
\left| g \right\ran \in {\cal H}_{/{\cal I} }\,
\Longleftrightarrow  
\forall f(z_1, z_2) \in {\cal I} , \quad
\left\lan 0,0 \right|
f^{\dagger}( \bar{z}_1, \bar{z}_2 ) 
\left| g \right\ran = 0 .
\ena
${\cal H}_{/{\cal I}}$ is a $k$-dimensional
space \cite{LecNakaj},
see appendix \ref{appNakaj}.

Let us denote the 
orthonormal complete basis of
${\cal H}_{/{\cal I}}$ by
$\left|\, g_{\a}  \, \right\ran , \a=1, 2,\cdots ,k$, 
and the 
complete basis of ${\cal H}_{{\cal I}}$ by 
$  \left|\, f_i \, \right\ran , i=k+1,k+2,\cdots $.
They span altogether the
complete basis of ${\cal H}$.
We can label them by positive integer $n$ : 
\bea
\{ \, \, \left|\, h_n \, \right\ran , 
\, \,  n \in {\bf Z}_+  \} =
\{ \, \, \left|\, g_\a \, \right\ran , 
\, \left|\, f_i \, \right\ran , 
\, \a = 1,2, \cdots , k ,\, \, 
 i = k+1, k+2, \cdots  \, \, \}.
\ena
As we can see from (\ref{inclusion}),
zero-modes (\ref{zeroV})
are completely
determined by  
the ideal $f_i(z_1,z_2)$
\cite{LecNakaj}:
\bea
\left| {\cal U}(f_i) \right\ran
=
\left( 
\begin{array}{c}
  \left| u_1(f_i) \right\ran \\
  \left| u_2(f_i) \right\ran \\
  \left|\, f_i \, \right\ran 
 \end{array}
\right) .
\ena
We can construct 
operator zero-mode (\ref{zeroPsi})
by the following formula:
\bea
 \label{genopzero}
\Psi = 
\sum_{i}\sum_{n}
(\Psi)_{in}
\left| {\cal U}(f_i) \right\ran \left\lan h_n \right| ,
\ena
where
$(\Psi)_{in}$ is a 
commuting number to be determined
by the normalization condition
of the zero-mode.
Let us first consider the
zero-mode which is normalized as
\bea
 \label{normpI}
\Psi_{0}^{\dagger} \Psi_{0} 
= p_{\cal I}  .
\ena
Using the gauge transformation
we can fix the form of such
zero-mode as follows:
\bea
 \label{nowind}
\Psi_{0} = 
\sum_{i}
(\Psi_0)_{i}
\left| {\cal U}(f_i) \right\ran 
\left\lan f_i \right| ,
\quad 
(\Psi_0)_{i} \ne 0 ,
\ena
If we write
\bea
 \label{Psi0}
\Psi_0 =
\left(
\begin{array}{c}
  (\psi_1)_0 \\
  (\psi_2)_0 \\
  \xi_0
\end{array}
\right) ,
\ena
(\ref{nowind}) means 
$\xi_0^{\dagger} = \xi_0$.
This means 
\bea
 \label{xi01}
\xi_0 \rightarrow 1 \quad 
(r \rightarrow \infty) .
\ena
As disscussed in the previous
subsection, 
the normalization (\ref{normpI}) causes
problem.
Thus just as in (\ref{MvNU}),
we need to find an operator
which satisfies
\bea
 \label{MvNPI}
U^{\dagger} U = \mbox{Id}_{\H}, \quad
U U^{\dagger} = p_{\cal I}.
\ena
Using the operator $U$,
we can construct the
zero-mode $\Psi$ which is
correctly normalized:
\bea
 \label{Psi0U}
 \Psi = \Psi_0 U^{\dagger}.
\ena
Then we obtain instanton solution
by (\ref{ncA}):
\bea
 \label{ncAideal}
A_\mu = \Psi^{\dagger} \pa_\mu \Psi.
\ena
From (\ref{xi01}) and (\ref{MvNPI})
we obtain the asymptotic behavior
of the gauge field (\ref{ncAideal}):
\bea
 \label{asAI}
A_\mu \rightarrow 
U \pa_\mu U^\dagger 
\quad (r \rightarrow \infty).
\ena
Thus the operator $U$
which satisfies eq.(\ref{MvNPI})
determins the asymptotic behavior
of the instanton gauge field,
and hence determins the
instanton number which reduces to the
surface integral.
In some simple instanton solutions,
we can explicitly show that the 
surface term essentially
reduces to the winding number
of gauge field around some $S^1$
at infinity, as we have done
in the previous subsection.
However, in general instanton solutions,
the expression of the
operator $U$ becomes complicated
and the relation to the
topological charge becomes
less clear. 
In the next section 
we give alternative 
explanation for the origin of the
integral instanton number.
We will show that the
instanton number is equal to the
dimension of the projection operator
$1 - p_{\cal I}$,
which is apparently integer.

In this subsection we have
learned that every $U(1)$ instanton
solution
on noncommutative ${\R}^4$ 
has corresponding projection operator to the ideal
states:
$p_{\cal I}$.

\subsection{Overlapping Instantons}

In this subsection we shall study the 
two-instanton solution
and observe what happens when
the two instantons approach and
overlap.
As we have learned in the previous subsection
the noncommutative $U(1)$
instanton solutions have
close relation with $p_{\cal I}$, the
projections to the ideal states.
Actually the correspondence 
between the noncommutative $U(1)$
instanton solutions with instanton number $k$
and ideal $\cal I \subset {\cal O}_{{\C}^2}$ with
$\mbox{dim}_{{\C}} \, {\cal O}_{{\C}^2}/{\cal I} = k$ is
one-to-one \cite{LecNakaj}, see also
appendix \ref{appNakaj}.
So it is enough to study
the behaviour of the $p_{\cal I}$, the 
projection to the ideal states 
that characterizes
the instanton solution. 

Let us consider the ideal
generated by 
$z_1$, $z_2$,
$z_1-w_1$ and  $z_2 - w_2$.
The corresponding 
projection operator to the ideal states
$p_{\cal I}$
projects out two dimensional
subspace spanned by
$\left|0,0 \right\ran$ 
and
$\left|\bw_1,\bw_2 \right\ran$.
Here 
$\left|\bw_1,\bw_2 \right\ran$
is the coherent state:
\bea
 \hbz_1 \left|\bw_1,\bw_2 \right\ran = 
 w_1 \left|\bw_1,\bw_2 \right\ran , \quad
  \hbz_2 \left|\bw_1,\bw_2 \right\ran = 
 w_2 \left|\bw_1,\bw_2 \right\ran.
\ena
It is related to $\left|0,0 \right\ran$
by  translation:
\bea
 \label{tcoh}
\left|\bw_1,\bw_2 \right\ran =
\exp [ \bw \cdot a^{\dagger} - w \cdot a ]
\left| 0, 0 \right\ran .
\ena
Here
\bea
\bw \cdot a^{\dagger} \equiv
\bw_1 a_1^{\dagger} + \bw_2 a_2^{\dagger},
\quad
w \cdot a \equiv
w_1 a_1 + w_2 a_2 \quad .
\ena
In order to construct a projection to the subspace
spanned by  $\left|0,0 \right\ran$ and
$\left|w_1,w_2 \right\ran$, 
it is convenient to construct
orthonormal basis.
We choose $\left|0,0 \right\ran$ 
for one basis vector and 
\bea
 \widetilde{\left| \bw_1, \bw_2 \right\ran}
\equiv
\frac{%
(1- \left| 0,0 \right\ran\left\lan 0,0 \right|)
\left|\bw_1, \bw_2 \right\ran %
}
{%
||(1- \left| 0,0 \right\ran\left\lan 0,0 \right|)
\left|\bw_1, \bw_2 \right\ran ||%
},
\ena
for another.
Then the projection operator can be written as
\bea
p_{\cal I}
= 1- \left| 0,0 \right\ran\left\lan 0,0 \right|
  - \widetilde{\left| \bw_1, \bw_2 \right\ran}
    \widetilde{\left\lan w_1, w_2 \right|}.
\ena
Now let us consider the limit
$|w|\equiv \sqrt{|w_1|^2 +|w_2|^2 } \rightarrow 0$.
From (\ref{tcoh}) we obtain
\bea
 \label{col}
\lim_{|w|\rightarrow 0}
\widetilde{\left| \bw_1, \bw_2 \right\ran}
= \frac{w\cdot a^{\dagger}}{|w|}
  \left| 0,0 \right\ran .
\ena
From (\ref{col}) we can observe that
the two-overlapping-instanton solution
has a collision angle  
$\frac{\vec{w}}{|w|}$ as a parameter,
besides its position.
However the collision angle has a redundancy
as a parameter of the solution
because phase of states
is not relevant to the projection.
Therefore the collision of two instantons
is parametrised by $S^3/U(1) \sim S^2$.
This describes
the $S^2$ that
appeared in the blowup
of singularity of the symmetric product of
${\R}^4$ \cite{Ours} from the noncommutative
points of view.  

\begin{figure}
\begin{center}
 \leavevmode
 \epsfxsize=130mm
 \epsfbox{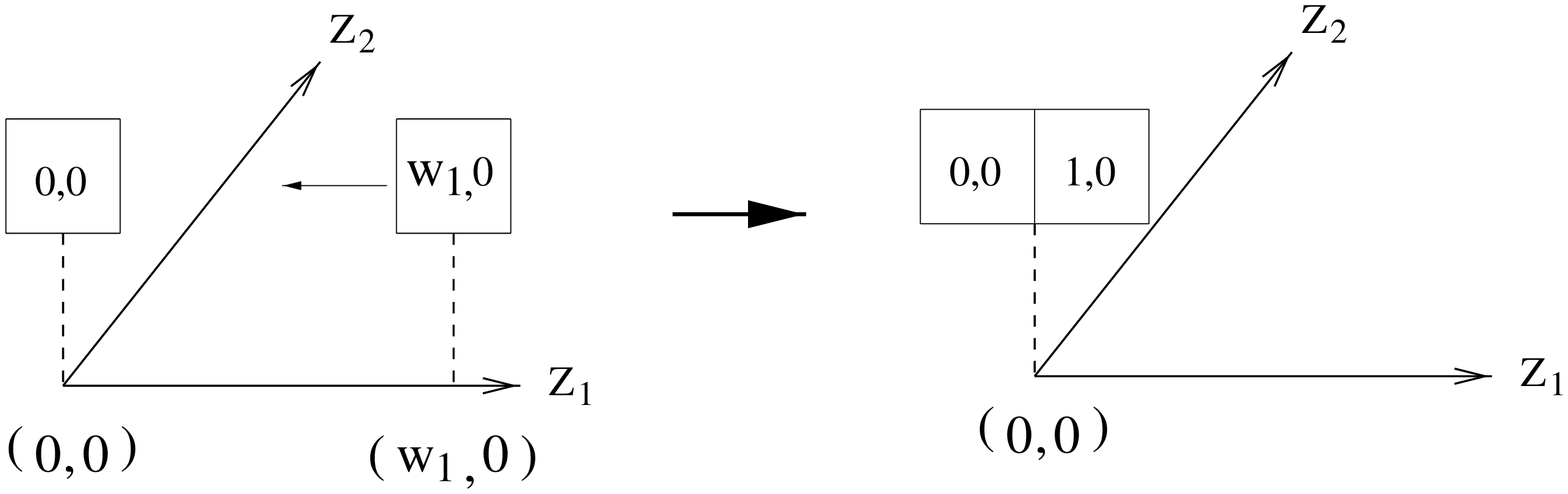}\\
\end{center}
\caption{
}\label{col1}
\end{figure}

\begin{figure}
\begin{center}
 \leavevmode
 \epsfxsize=130mm
 \epsfbox{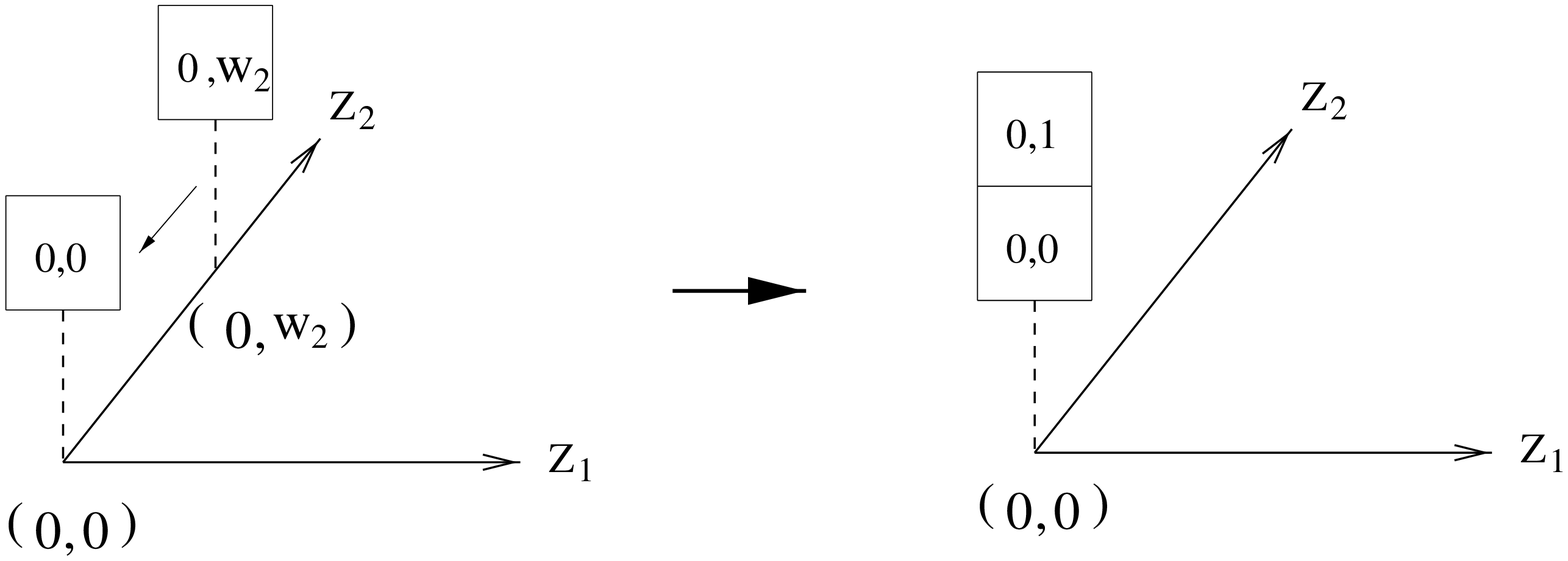}\\
\end{center}
\caption{
}\label{col2}
\end{figure}

Fig.\ref{col1} and fig.\ref{col2}
schematically show  
two typical overlapping processes.
The boxes express the projections to the states.
\framebox[8mm]{\rule{0pt}{4mm}\footnotesize 0,0}
represents the projection to the
Fock vacuum 
$\left|0,0 \right\ran$,
\framebox[8mm]%
{\rule{0pt}{4mm}\footnotesize $w_1$,0}
represents the projection to the
coherent state
$\left|\bw_1,0 \right\ran$,
\framebox[8mm]{\rule{0pt}{4mm}\footnotesize 1,0}
represents the projection to
the first excited states  
$\left| 1,0 \right\ran$, etc.\footnote{%
The notation is little bit
confusing because the expressions for
the coherent states
with $(\bw_1,\bw_2) = (1,0)$ and
the first excited states in occupation number
representation $(n_1,n_2) = (1,0)$ coincide.
Here by
$\left| 1,0 \right\ran$ we mean
the first excited states.}
Recall that the
Wick symbol of the projection operator to
the coherent states
$\left|\bw_1,\bw_2 \right\ran$
concentrates around 
$(z_1,z_2)=(w_1,w_2) $.
Fig.\ref{col1} shows the
case $\frac{\vec{w}}{|w|}
= (1,0)$ and
fig.\ref{col2}
shows the
case $\frac{\vec{w}}{|w|}
= (0,1)$.
It is interesting to observe that
although both
$\left| 0,0 \right\ran$
and
$\left| \bw_1, \bw_2 \right\ran$
are (the translation of)
the zero-th excited states in the
Fock space,
after the overlap
$(|w| \rightarrow 0)$
the first excited states like
$\left| 1,0 \right\ran$ or
$\left| 0,1 \right\ran$
appear.

\newpage
\section{Partial Isometry in IIB Matrix Model %
and \\
Instanton Number as %
Dimension of Projection}\label{secIIB}

In section \ref{oneone} we have studied
how the effect of noncommutativity gives
non-zero instanton number to the 
$U(1)$ gauge field.
In the case of
$U(1)$ one-instanton solution
the effect of noncommutativity
does not vanish on all the $S^3$ at infinity,
but the gauge field concentrates
around $S^1$.
The instanton number
essentially reduces to the winding
number of gauge field around this $S^1$.
However it was not quite
clear why the instanton number
correctly reduces to
the winding number including the
numerical coefficient.
It was also difficult to apply the
arguments
for the general solutions.
On the other hand,
as we have learned in section \ref{tideal},
noncommutative
$U(1)$ instantons have close relation with the
projection operators. 
Therefore
it is desirable to have an 
explanation of the instanton number
directly related to the projection
operator.
For that purpose it is necessary to
introduce the notion of the
covariant derivative which acts only
in the subspace of the Fock space 
\cite{mine}\cite{mine2}.
In order to treat such generalized gauge theory,
it is convenient to treat 
noncommutative Yang-Mills theory
in the framework of IIB matrix model \cite{IKKT}.
In IIB matrix model, noncommutative
Yang-Mills theory appears
from the expansion around certain
background \cite{IKKTB}.

IIB matrix model was
proposed as a non-perturbative formulation of
type IIB superstring theory \cite{IKKT}.
It is defined by the following action:
\bea
 \label{IIB}
S = -\frac{1}{g^2}
\mbox{Tr}_{U(N)}
\left(
\frac{1}{4}[X_\mu, X_\nu][X^\mu, X^\nu]
+
\frac{1}{2}\bar{\Theta}\Gamma^\mu
[X_\mu,\Theta]
\right)\quad (\mu = 0,\cdots,9),
\ena
where 
$X_\mu$ and $\Theta$ are $N \times N$
hermitian matrices and each component
of $\Theta$ is a Majorana-Weyl spinor.
The action (\ref{IIB}) 
has the following $U(N)$ symmetry:
\bea
 \label{UNS}
X_\mu &\rightarrow & UX_\mu U^{\dagger}, \nn
\Theta& \rightarrow & U\Theta U^{\dagger},
\ena
where $U$ is an $N\times N$ unitary matrix:
\bea
UU^{\dagger} = U^{\dagger}U = \mbox{Id}_N.
\ena
The action (\ref{IIB}) also
has the following ${\cal N}=2$ 
supersymmetry:
\bea
\label{IISUSY}
\delta^{(1)} \Theta 
&=& \frac{i}{2} [X_{\mu},X_{\nu}] 
\Gamma^{\mu\nu} \epsilon^{(1)}, \nn  
\delta^{(1)} X_\mu 
&=& i\bar{\epsilon}^{(1)} \Gamma_{\mu} \Theta , \nn
\delta^{(2)} \Theta &=& \epsilon^{(2)}, \nn
\delta^{(2)} X_\mu &=& 0.
\ena

Noncommutative
Yang-Mills theory appears
when we consider
the model in certain classical
background \cite{IKKTB}.
This background
is a  solution to the classical equation of motion
and identified with 
D-brane in type IIB superstring theory.
The classical equation of motion 
of IIB matrix model is 
given by
\bea
 \label{IIBeq}
[X_{\mu},[X_{\mu},X_{\nu}]] =0.
\ena 
One class of solutions
to (\ref{IIBeq}) is given
by simultaneously diagonalizable
matrices, i.e.
$[X_{\mu},X_{\nu}] = 0$ for all $\mu,\nu$.
However
IIB matrix model has another class
of
classical solutions which are
interpreted as
D-branes in type IIB superstring theory:
\bea
 \label{half}
& &X_\mu = i \hat{\pa}_\mu, \nn
& &[i\hat{\pa}_\mu, i\hat{\pa}_\nu] = - i B_{\mu\nu},
\ena
where 
$B_{\mu\nu}$ is a 
constant matrix.
$i\hat{\pa}_\mu$'s are
infinite rank matrices
because if they have only finite rank,
taking a trace
of both sides of
(\ref{half}) results in an
apparent 
contradiction.
(\ref{half}) is the same as
the one appeared in section \ref{ncR},
(\ref{comdel}).
Therefore we define 
``coordinate matrices" $\hat{x}^{\mu}$ from
the formula (\ref{deri}):
\bea
\hat{x}^{\mu} \equiv 
- i \theta^{\mu \nu}\hat{\pa}_{\nu}.
\ena
Then their commutation relations take the
same form as in (\ref{noncomxmn}): 
\bea
 \label{xxB}
[\hat{x}^{\mu}, \hat{x}^{\nu} ] = i \theta^{\mu\nu},
\ena
where $\theta^{\mu\nu}$ is an inverse
matrix of $B_{\mu\nu}$.
We identify these infinite 
dimensional matrices with operators
acting in Fock space $\H$
which have been discussed in the
previous sections.
Thus 
noncommutative coordinates of
${\R}^{2d}$
appear as a classical
solution of
IIB matrix model,
where $2d$ is rank of $B_{\mu\nu}$
and dimension of noncommutative
directions.
Now let us expand fields around
this background:
\bea
X_{\mu} &=& i (\hat{\pa}_{\mu} + A_{\mu})
        \equiv i \hat{D}_\mu , \label{bgf} \\
X_{I} &=& \Phi_I . \label{bgphi}
\ena
Here $\mu , \nu $ are the indices of the noncommutative 
directions, i.e. $det\, \theta^{\mu\nu} \ne 0$ and
$I,J$ is the indices of the directions transverse to 
the noncommutative directions.
We shall call $\hat{D}_\mu$
{\bf covariant derivative operator}.
Then the action (\ref{IIB}) becomes
\bea
S &=&  -\frac{1}{g^2}
\mbox{Tr}_{\H} \Biggl[
- \frac{1}{4}
(F_{\mu\nu}+iB_{\mu\nu})
(F^{\mu\nu}+iB^{\mu\nu})
+
\frac{1}{2}
D_\mu \Phi_I D^{\mu} \Phi_I \Biggr. \nn
& &\qquad \qquad \quad
\Biggl. +
\frac{1}{4}
[\Phi_I, \Phi_J][\Phi_I, \Phi_J]
+
\frac{1}{2}\bar{\Theta}\Gamma^\mu
D_\mu \Theta
+
\frac{1}{2}\bar{\Theta}\Gamma^I
[\Phi_I,\Theta] \Biggr].
\ena
Here
\bea
D_\mu \Phi_I 
&\equiv&
[\hat{D}_\mu , \Phi_I]
=
\pa_\mu \Phi_I + [A_\mu, \Phi_I ], \\
D_\mu \Theta 
&\equiv&
[\hat{D}_\mu ,\Theta]
=
\pa_\mu \Theta_I + [A_\mu, \Theta_I ] .
\ena
Hence we obtain supersymmetric noncommutative
$U(1)$ gauge theory. 
The gauge transformation
follows from (\ref{UNS}):
\bea
A_\mu &\rightarrow & UA_\mu U^{\dagger} 
+ U\pa_\mu U^{\dagger}, \label{GA}\\
\Phi_I &\rightarrow & U\Phi_I U^{\dagger}, \\
\Theta &\rightarrow & U\Theta U^{\dagger} .
\ena
Here $U$ is a unitary operator:
\bea
 U U^{\dagger} = U^{\dagger} U = \mbox{Id}_{\H}.
\ena
The transformation of the gauge field (\ref{GA})
is determined by the requirement
that the form of the derivative of operators
should be kept
under the gauge transformation.
As described in
section \ref{ncR},
we can rewrite above matrix multiplication
using ordinary functions and star products.

\subsection{Partial Isometry}

In section \ref{oneone} we have encountered with an  
operator $U$ which satisfies the relation 
like\footnote{The role of $U$ and $U^{\dagger}$ here 
are
exchanged from section \ref{oneone}. See
the generalised framework in this section.
}
\bea
 \label{iso}
U U^{\dagger} = p, \quad
U^{\dagger}U  = \mbox{Id}_{\H},
\ena
where $p$ is some projection: $p=p^{\dagger}, p^2 =p$.
Let us examine the physical meaning of such 
operators in IIB matrix model.
$U$ gives a one-to-one map between
$\H$ and $p \H$. Accordingly
the operator $U$ gives a map for 
all the fields in IIB matrix model:
\bea
 \label{mvn1}
X^{(p)}_\mu \equiv UX_\mu U^{\dagger}, \quad
\Theta^{(p)} \equiv U \Theta U^{\dagger}.
\ena
Here the
action of the operators with subscripts $(p)$
are restricted to $p\H$.
This map is mere a change of the names
of the states.
For example, let us consider a simple
example of such operator in noncommutative
${\R}^2$.
\bea
 \label{simpleiso}
U = \sum_{n=0}^{\infty} 
   \left| n +1 \right\ran
 \left\lan  n  \right|,
\ena
\bea
U U^{\dagger} = p 
= 1-\left| 0 \right\ran
 \left\lan  0  \right|, \quad
U^{\dagger}U  = \mbox{Id}_{\H} \quad .
\ena
Under this map $U$ the state which was called
$\left| n  \right\ran $ is renamed
to $\left| n +1 \right\ran$.
The change of names should not change
physics. 
In this regard this map may be regarded
as an noncommutative analog of 
coordinate transformation.
Hence
$X^{(p)}_\mu$ satisfies equation of motion (\ref{IIBeq})
if $X_\mu$ does:
\bea
[X^{(p)}_\mu, [ X^{(p)}_\mu , X^{(p)}_\nu ] ]=
U [X_\mu, [X_\mu , X_\nu] ] U^{\dagger} = 0. 
\ena
However note that
this change of names
causes a change in the expressions
in operator symbols.
The covariant derivative operator 
$\hat{D}_\mu^{(p)}$
for the restricted subspace $p\H$
is defined by
\bea
 \label{pDp}
\hat{D}_\mu^{(p)} \equiv
U \hat{D}_\mu  U^{\dagger} .
\ena
We separate the
derivative part and 
gauge field 
part as follows:
\bea
 \label{pDp2}
\hat{D}_\mu^{(p)}
\equiv
p\hat{\pa}_\mu p + A_\mu^{(p)} .
\ena
It is appropriate to regard 
$p\hat{\pa}_\mu p$ as a derivative
operator for $p\H$ since
for the operator
restricted to $p\H$; 
$O^{(p)} = pO^{(p)}p $,
the action of derivative
is equal to
the commutator with $p\hat{\pa}_\mu p$:
\bea
\pa_\mu O^{(p)} \equiv
[\hat{\pa}_\mu, O^{(p)}] = 
[p\hat{\pa}_\mu p , O^{(p)}].
\ena
From eq.(\ref{pDp}), we determine
the relation between
$A_\mu$ and $A_\mu^{(p)}$:
\bea
\hat{D}_\mu^{(p)} =
U \hat{D}_\mu U^{\dagger}
&=&
U \hat{\pa}_\mu U^{\dagger} + 
 U A_\mu U^{\dagger}
=
U \hat{\pa}_\mu U^{\dagger} p + 
 U A_\mu U^{\dagger} \nn
&=&
U [\hat{\pa}_\mu, U^{\dagger}] p
+ p \hat{\pa}_\mu p + U A_\mu U^{\dagger}.
\ena
Thus we obtain the relation
\bea
A_\mu^{(p)} 
= U A_\mu U^{\dagger} +
U \pa_\mu  U^{\dagger} p .
\ena
Next we determine the 
transformation rule
of the field strength.
The commutator of the covariant derivative
operators transform as
\bea
 \label{MvNF}
F_{\mu\nu} + i B_{\mu\nu}
=
[\hat{D}_\mu , \hat{D}_\nu] 
\rightarrow 
[\hat{D}_\mu^{(p)} , \hat{D}_\nu^{(p)}]
&=& U[\hat{D}_\mu , \hat{D}_\nu]U^{\dagger}
= U(F_{\mu\nu} + i B_{\mu\nu} ) U^{\dagger} \nn
&=&
UF_{\mu\nu}  U^{\dagger}+ i p B_{\mu\nu} .
\ena
From (\ref{MvNF})
we determine the 
transformation
of field strength as follows:
\bea
F_{\mu\nu} 
&\rightarrow&
F_{\mu\nu}^{(p)} \equiv UF_{\mu\nu}  U^{\dagger}
\label{MvNF2} \\
& &\qquad =
[\hat{D}_\mu^{(p)} , \hat{D}_\nu^{(p)}] - i p B_{\mu\nu}
\label{-pB}
\ena
This means that for the field strength
a map $U$ is mere a change of names of states
as in (\ref{mvn1}).
As a consequence  instanton number
does not change under
this transformation:
\bea
 \label{MvNI}
-\frac{1}{16 \pi^2} \int d^4x 
F_{\mu\nu} \tilde{F}^{\mu\nu}
=
-\frac{1}{16 \pi^2} \int d^4x 
F_{\mu\nu}^{(p)} \tilde{F}^{(p)\mu\nu}.
\ena
Here 
$\tilde{F}_{\mu\nu}
\equiv \frac{1}{2} {\epsilon_{\mu\nu}}^{\rho\sigma}
F_{\rho\sigma}$ .
However while the left hand side of (\ref{MvNI})
can be rewritten as surface term,
the right hand side cannot.
This is because 
the derivative part in $p\H$is different
from the one in $\H$, see (\ref{pDp2}).

It will become crucial 
in the calculation of instanton number
that the last term in eq.(\ref{-pB})
is proportional to the
projection operator $p$.

Now let us consider more general situation.
Consider operator $U$ which satisfies
\bea
 \label{MvNpq}
p = U^{\dagger} U \quad  \mbox{and} \quad 
q = U U^{\dagger} .
\ena
Here both $p$ and $q$ are projection.
The two projections  $p$ and $q$ 
are said to be equivalent,
or {\bf Murray-von Neumann equivalent}
when there exists $U$ 
which satisfies (\ref{MvNpq})
and written
as $p \sim q$.\footnote{%
For a detail of these projection calculus,
see for example
\cite{WO}.}
$U$ is called {\bf partial isometry} when
$U^{\dagger}U$ is a projection.
If $U^{\dagger}U = \mbox{Id}_\H$,
$U$ is said to be {\bf isometry}. 
Following are the fundamental properties of 
the partial isometry:
\bea
U = Up = q\, U ,\quad U^{\dagger} = p \, 
U^{\dagger} = U^{\dagger}q .
\ena
\bea
\mbox{ker} \, U 
= \mbox{Id}_{\H} - U^{\dagger} U , \quad
\mbox{ker} \, U^{\dagger} 
=  \mbox{Id}_{\H} - U U^{\dagger}.
\ena
\bea
 \label{up}
U^{\dagger} \H 
= U^{\dagger}U \H = p \H, \quad
U\H = UU^{\dagger} \H = q \H  .
\ena
By choosing 
orthonormal basis of $p \H$ and
$q \H$, it is easily shown that
\bea
 \label{pHqH}
p \sim q  \Leftrightarrow
\mbox{dim}\, p \H = \mbox{dim} \, q \H.
\ena
Note that
$p$ can be equivalent to the identity
if $p$ has infinite rank, for example
see eq.(\ref{simpleiso}).

Under the map from $p\H$ to $q\H$
by $U$ which satisfies (\ref{MvNpq}),
operators transform as
\bea
 \label{mvn}
X^{(q)}_\mu = UX_\mu^{(p)} U^{\dagger}, \quad
\Theta^{(q)} = U \Theta^{(p)} U^{\dagger}.
\ena
The transformation rule of 
gauge field under 
partial isometry
is determined from (\ref{mvn})
\bea
\hat{D}_\mu^{(p)} \rightarrow
\hat{D}_\mu^{(q)} 
&=&
U \hat{D}_\mu^{(p)} U^{\dagger}
=
U p\hat{\pa}_\mu p U^{\dagger}
+ U A_\mu^{(p)} U^{\dagger} \nn
&=&
q \hat{\pa}_\mu q 
+ U [\hat{\pa}_\mu, U^{\dagger}] q 
+ U A_\mu^{(p)} U^{\dagger} .
\ena
Hence we obtain the transformation of the
gauge field under the map $U$:
\bea
 \label{MvNgt}
A_\mu^{(p)} \rightarrow
A_\mu^{(q)} \equiv
U \pa_\mu U^{\dagger} q + U A_\mu^{(p)} U^{\dagger}.
\ena
Since the form of the transformation
is similar to the gauge transformation
(\ref{gauget}),
we will call (\ref{MvNgt})
{\bf Murray-von Neumann (MvN) 
gauge transformation}.\footnote{%
The idea of extending the notion of gauge
transformation by using the non-unitary operator
which satisfies (\ref{iso})
was first appeared in \cite{Ho}. }
Indeed, MvN gauge transformation contains
noncommutative counterparts of the 
singular gauge transformation in
commutative space, as we will
observe in the following
and  in appendix \ref{appU2}. 
Under the MvN gauge transformation
the field strength transforms as
\bea
F_{\mu\nu}^{(p)} \rightarrow
F_{\mu\nu}^{(q)} \equiv 
U F_{\mu\nu}^{(p)} U^{\dagger} .
\ena

\subsection{ADHM Construction 
within the Projected Space}\label{padhm}

Here we show that the
gauge field constructed from
zero-mode given in 
(\ref{nowind}):
\bea
 \label{Apasd}
A_{\mu}^{({p_{\cal I}})} 
\equiv
\Psi_0^{\dagger}[\hat{\pa}_{\mu},\Psi_0] {p_{\cal I}},
\ena
is anti-self-dual
provided that the field strength is
constructed in $\H_{\cal I} = {p_{\cal I}} \H$
as in (\ref{MvNF2}).
Actually (\ref{Apasd}) is a
MvN transform of
the gauge field constructed
from the zero-mode normalized as 
in (\ref{norm}).
Then since MvN gauge transformation 
does not affect the Lorentz indices,
anti-self-duality of $A_{\mu}^{({p_{\cal I}})} $
is a straightforward consequence.
But it is also easy to check the
anti-self-duality of $A_{\mu}^{({p_{\cal I}})} $
directly.  
We treat the gauge field $A_{\mu}^{({p_{\cal I}})} $
within the framework of IIB matrix model:
\bea
X_{\mu} 
&=& i\hat{D}_\mu = 
 {p_{\cal I}}(i\hat{\pa}_{\mu} + i A_{\mu}){p_{\cal I}} 
\quad (\mu = 1,\cdots, 4) \nn
&=& {p_{\cal I}}(i\hat{\pa}_{\mu}){p_{\cal I}} 
+ {p_{\cal I}}(i \Psi_0^{\dagger} \hat{\pa}_\mu \Psi_0 )
     {p_{\cal I}} -
{p_{\cal I}}(i \Psi_0^{\dagger} \Psi_0 {p_{\cal I}} 
     \hat{\pa}_{\mu}){p_{\cal I}} \nn
&=& i {p_{\cal I}} \Psi_0^{\dagger} \hat{\pa}_\mu \Psi_0 
        {p_{\cal I}} 
= i \Psi_0^{\dagger} \hat{\pa}_\mu \Psi_0 .
\label{IIBD}
\ena
From (\ref{IIBD}) we obtain
\bea
 \label{XX}
[X_{\mu}, X_{\nu}]
=
{p_{\cal I}}(-i B_{\mu\nu} 
- F^{(p_{\cal I})-}_{\mu\nu\, \mbox{\tiny ADHM} })
{p_{\cal I}} \quad .
\ena
and 
$F_{\mu\nu}^{({p_{\cal I}})} 
= [\hat{D}^{(p_{\cal I})}_\mu,\hat{D}^{(p_{\cal I})}_\nu] -
i{p_{\cal I}}B_{\mu\nu} =
 F^{({p_{\cal I}})-}_{\mu\nu\, \mbox{\tiny ADHM} } $ 
is anti-self-dual. 
Indeed,
\bea
[X_{\mu}, X_{\nu}]
&=& 
i {p_{\cal I}} \Psi_0^{\dagger} \hat{\pa}_\mu 
   \Psi_0 {p_{\cal I}} \cdot 
i p \Psi_0^{\dagger} \hat{\pa}_\nu \Psi_0 {p_{\cal I}} 
-(\mu \leftrightarrow \nu) \nn
&=&
- {p_{\cal I}} \Psi_0^{\dagger} 
 \hat{\pa}_\mu \Psi_0 \Psi_0^{\dagger} 
 \hat{\pa}_\nu \Psi_0 {p_{\cal I}} 
-(\mu \leftrightarrow \nu) \nn
&=&
\left\{
{p_{\cal I}} \Psi_0^{\dagger} 
 \hat{\pa}_\mu (1-\Psi_0\Psi_0^{\dagger}) 
\hat{\pa}_\nu \Psi_0 {p_{\cal I}} 
-(\mu \leftrightarrow \nu) \right\}
- {p_{\cal I}}\Psi_0^{\dagger}  [\hat{\pa}_\mu , \hat{\pa}_\nu ]
  \Psi_0  {p_{\cal I}} \nn
&=&
\left\{{p_{\cal I}} \Psi_0^{\dagger} \hat{\pa}_\mu 
{\cal D}_z^{\dagger}
\frac{1}{ {\cal D}_z {\cal D}_z^{\dagger}  }
{\cal D}_z  
\hat{\pa}_\nu \Psi_0 {p_{\cal I}} 
 -(\mu \leftrightarrow \nu) \right\}
- {p_{\cal I}}(iB_{\mu\nu}){p_{\cal I}} \quad .
\ena
Remaining calculations are the same as the one in
(\ref{cFS}) and we obtain
\bea
 \label{PFS}
F^{(p_{\cal I})}
&=&
\Psi_0^{\dagger }
\left(
 \begin{array}{ccc}
dz_1 \frac{1}{\, \, \Box_z}d\bar{z}_1 
+ d\bar{z}_2 \frac{1}{\, \, \Box_z} dz_2   & 
  -dz_1 \frac{1}{\, \, \Box_z} d\bar{z}_2 
       + d\bar{z}_2\frac{1}{\, \, \Box_z} 
dz_1 & 0 \\
-dz_2 \frac{1}{\, \, \Box_z} d\bar{z}_1 
+ d\bar{z}_1\frac{1}{\, \, \Box_z} dz_2 & 
  dz_2 \frac{1}{\, \, \Box_z} d\bar{z}_2 
  + d\bar{z}_1 \frac{1}{\, \, \Box_z} dz_1 & 0 \\
 0 & 0 & 0
 \end{array}
\right) 
\Psi_0 \nn
&\equiv & F^{(p_{\cal I})-}_{\mbox{\tiny ADHM}} .
\ena

\subsection{Instanton Number %
as Dimension of Projection}\label{inump}

In the following we shall prove
that the instanton number 
of the gauge field $A_\mu^{({p_{\cal I}})} $
in (\ref{Apasd})
is equal to the dimension of the
projection $(1-p_{\cal I})$.

The field strength constructed from
(\ref{Apasd}) is anti-self-dual provided that
it is defined in
the restricted subspace $\H_{\cal I}$:
\bea
 \label{Fp}
F_{\mu\nu}^{({p_{\cal I}})}
= 
[\hat{D}_\mu^{({p_{\cal I}})}, \hat{D}_\nu^{({p_{\cal I}})} ]- 
  i {p_{\cal I}} B_{\mu\nu} \quad .
\ena
The fact that the
last term in (\ref{Fp}) is proportional to
the projection operator $p_{\cal I}$
expresses that this field strength is 
constructed in ${\H_{\cal I}}$.
Now let us regard the {\it same}
covariant derivative operator 
as covariant derivative operator
for the {\em full} Fock space $\H$:
\bea
{\hat{D}'}_\mu = \hat{\pa}_\mu + {A'}_\mu
=
\hat{D}^{(p)}_\mu
=
{p_{\cal I}}\hat{\pa}_\mu {p_{\cal I}} 
+ A_\mu^{({p_{\cal I}})} \quad .
\ena
Here 
${A'}$ is {\it not} an MvN gauge transform 
of $A_{p_{\cal I}}$ but
\bea
 {A'}_{\mu} &=&
{p_{\cal I}}\hat{\pa}_{\mu}{p_{\cal I}}
- \hat{\pa}_{\mu} + A_{\mu}^{({p_{\cal I}})}   \nn
&=&
-(1-{p_{\cal I}})\hat{\pa}_{\mu}-\hat{\pa}_{\mu}
  (1-{p_{\cal I}})
+(1-  {p_{\cal I}} )\hat{\pa}_{\mu}(1-{p_{\cal I}}) 
  + A_{\mu}^{({p_{\cal I}})}      \quad .
\ena
The field strength of ${A'}_{\mu}$
is defined in the full Fock space $\H$:
\bea
 \label{pFp}
{F'}_{\mu\nu} = 
[\hat{D}_\mu^{({p_{\cal I}})}, 
\hat{D}_\nu^{({p_{\cal I}})} ]-  iB_{\mu\nu}\quad .
\ena
The last term in (\ref{pFp}) expresses that
this is a field strength in the
full Fock space $\H$.
Although the covariant derivative
operators are the same,
the difference between the field strengths
${F'}_{\mu\nu}$ and $F_{\mu\nu}^{({p_{\cal I}})}$
arises from the last terms in
(\ref{Fp}) and (\ref{pFp}) 
by definition (\ref{-pB}):
\bea
{F'}_{\mu\nu} = 
F_{\mu\nu}^{({p_{\cal I}})} -  (1-{p_{\cal I}}) iB_{\mu\nu}
\ena
Since the gauge field  $A'$ is the
gauge field for the full Fock space,
the instanton number of $A'$ can be rewritten
as a surface term in the same way
as in the commutative case:
\bea
 \label{sfc}
-\frac{1}{16 \pi^2}
\left(2\pi\right)^2 \sqrt{ det \theta}\,
\mbox{Tr}_{\H}
{F'}_{\mu\nu} {\tilde{F}}^{'\mu\nu} 
&=&
-\frac{1}{16 \pi^2}
\int d^4x\,
{F'}_{\mu\nu} \star_N {\tilde{F}}^{'\mu\nu} \nn
&=&
-\frac{1}{16 \pi^2}
\int d^4x\,
\pa_{\mu} K^{\mu}.
\ena
Here 
$K^{\mu}$ is defined as
\bea
K^{\mu} = 
2\, 
\epsilon^{\mu\nu\rho\sigma}
\left(
A'_{\nu}\star_N \pa_{\rho} A'_{\sigma} 
+ \frac{2}{3} A'_{\nu}\star_N A'_{\rho} \star_N A'_{\sigma}
\right).
\ena
However the instanton number
of $A_\mu^{({p_{\cal I}})}$ cannot be written as
a surface term, because the derivative part
for the subspace $\H_{\cal I}$ 
is different from that for $\H$,
see (\ref{pDp2}). 
From the form of $\xi_0$ in
(\ref{Psi0}),
the instanton number (\ref{sfc})
of ${A'}_\mu$ vanishes,
as described in section \ref{ncADHM}.
On the other hand,
\begin{eqnarray}
0 &=&\, \frac{1}{16 \pi^2}
\left(2\pi\right)^2 \sqrt{ det \theta}\, \,
\mbox{Tr}_{\H}\, \,
{F'}_{\mu\nu} {F}^{'\mu\nu} \nn
&=&\, 
\frac{1}{16 \pi^2}\left(2\pi\right)^2
\sqrt{det \theta}\, \, \mbox{Tr}_{\H}\,
\left[
(1-{p_{\cal I}})B_{\mu\nu}B^{\mu\nu} -
 F_{\mu\nu}^{({p_{\cal I}})} F^{({p_{\cal I}})\mu\nu}
\right],
\end{eqnarray}
(here $B_{\mu\nu}$ is anti-self-dual
as we have set $\theta^{\mu\nu}$ anti-self-dual).
Thus the instanton number counts the dimension of the
projection $1 - p_{\cal I}$ :
\bea
-
\frac{1}{16 \pi^2}
\left(2\pi\right)^2 \sqrt{ det \theta}\,
\mbox{Tr}_{\H}\,
F_{\mu\nu}^{({p_{\cal I}})} 
\tilde{F}^{({p_{\cal I}})\, \mu\nu}  
&=&
\frac{1}{16 \pi^2}
\left(2\pi\right)^2 \sqrt{ det \theta}\,
\mbox{Tr}_{\H}\,
F_{\mu\nu}^{({p_{\cal I}})} F^{({p_{\cal I}})\, \mu\nu}  \nn
&=&\, 
\frac{1}{16 \pi^2}
\left(2\pi\right)^2 \sqrt{ det \theta}\,
\mbox{Tr}_{\H}\,
(1-{p_{\cal I}})B_{\mu\nu}B^{\mu\nu} \nn
&=&\,  \mbox{dim}\, (1-{p_{\cal I}}).
\end{eqnarray}
Recall that the
instanton number is invariant
under the MvN gauge transformation,
see eq.(\ref{MvNI}).
This means
\bea
(\mbox{winding number}) + 
(\mbox{dimension of the subspace projected out}).
\ena
is conserved under the MvN gauge transformation.

In the above we have used the
self-duality of $B_{\mu\nu}$,
but this condition is not essential.
The proof for the case when $B_{\mu\nu}$ 
is not self-dual
goes much in the same way.

The origin of the integral instanton number
is naturally understood from following
relation, see eqs.(\ref{op-fn}):
\bea
\frac{1}{16\pi^2}
\int d^4x B_{\mu\nu} \tilde{B}^{\mu\nu}
=
\frac{1}{16\pi^2}
(2 \pi)^2 \sqrt{det \theta}\, 
\mbox{Tr}_\H B_{\mu\nu} \tilde{B}^{\mu\nu}
=\mbox{Tr}_\H  \quad .
\ena

\subsection{Unification of Gauge Field and Geometry}

\subsubsection*{Example: $U(1)$ One-Instanton Again}

\begin{figure}
\begin{center}
 \leavevmode
 \epsfxsize=120mm
 \epsfbox{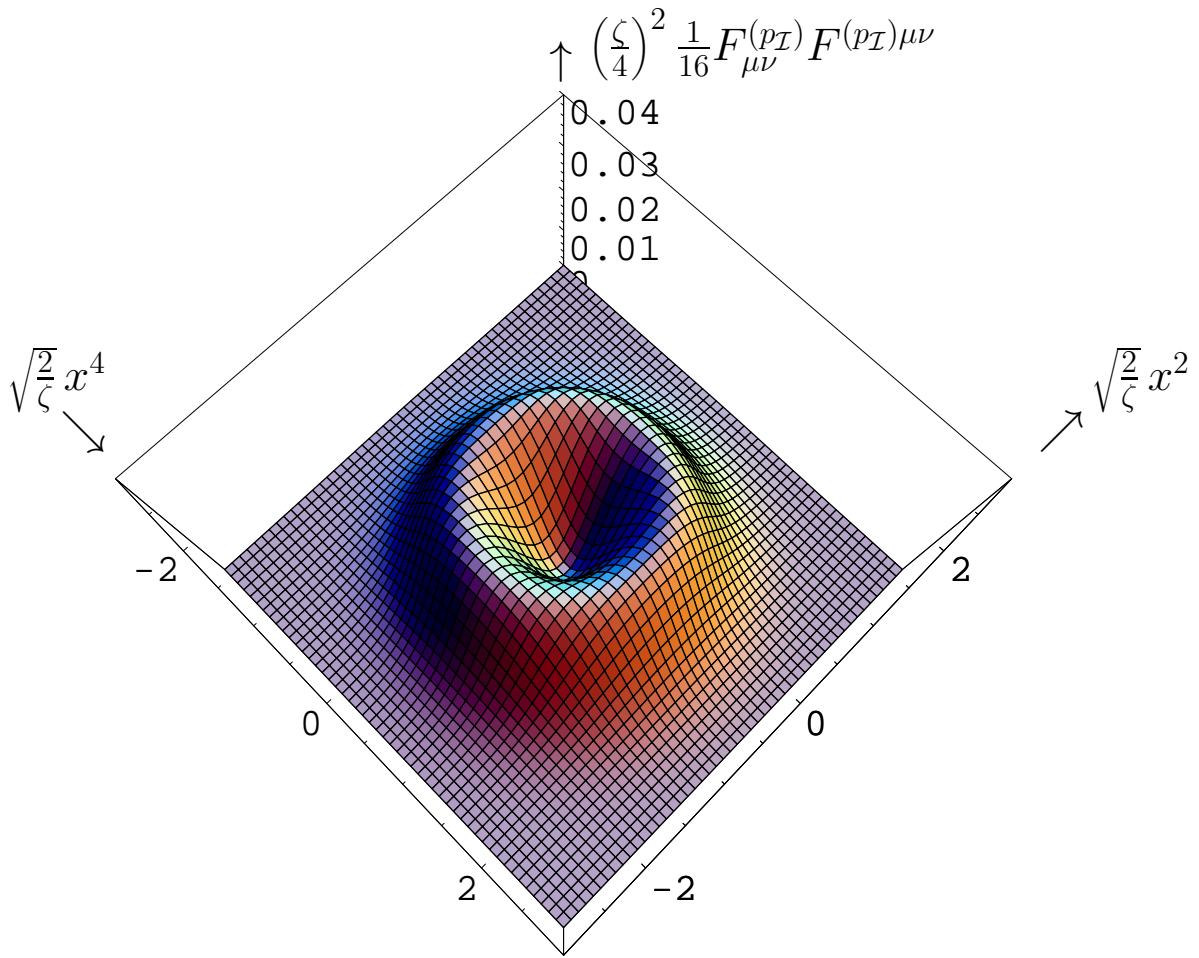}\\
\begin{picture}(0,0)
\put(180,210){\Large $\nearrow$}
\put(200,230){\Large $\sqrt{\frac{2}{\zeta}}\, x^2$}
\put(-210,230){\Large $\sqrt{\frac{2}{\zeta}}\, x^4$}
\put(-190,210){\Large $\searrow$}
\put(-5,350){\Large $\uparrow $}
\put(10,355){\Large $\left(\frac{\zeta}{4} \right)^2
\frac{1}{16}
F_{\mu\nu}^{(p_{\cal I})}F^{(p_{\cal I})\mu\nu}$}
\end{picture}
\caption{$x^1=x^3=0$ slice of the Wick symbol of 
$\left(\frac{\zeta}{4} \right)^2
\frac{1}{16}F_{\mu\nu}F^{\mu\nu}$. Observe
that it vanishes at the origin of ${\R}^4$. 
}\label{ffone3D} 
\end{center}
\end{figure}

Let us study
the physical meaning 
of the MvN gauge transformation
more closely by
taking the $U(1)$ one-instanton solution
as an example.
$U(1)$ one-instanton solution is 
given in (\ref{Ainst}):  
\bea
 \label{Ainst2}
A_\mu = \Psi^{\dagger} \pa_\mu \Psi ,
\ena
where (see (\ref{FPsi0}) $\sim$ (\ref{PoU}) )
\begin{eqnarray}
 \label{PoU2}
 \Psi = \Psi_0 U_1^{\dagger}\quad .
\end{eqnarray}
We can MvN gauge transform 
$A_\mu$ by $U_1$:
\bea
A_{\mu}^{p_{\cal I}} =
U_1^{\dagger} \pa_\mu U_1 {p_{\cal I}}
+ U_1^{\dagger}  A_\mu U_1 ,
\ena
where ((\ref{MvNU}))
\begin{eqnarray}
 U_1U^{\dagger}_1 = 1, \quad 
 U_1^{\dagger}U_1 = p_{\cal I} =
1 - 
\left| 0,0 \right\ran
\left\lan 0,0\right|   .
\end{eqnarray}
It is easy to show (see also (\ref{Apasd}))
\bea
A_{\mu}^{p_{\cal I}} =
\Psi^{\dagger}_0 ( \pa_\mu \Psi_0 )  {p_{\cal I}} .
\ena
This MvN gauge transformation $U_1$
unwind the winding of gauge field $A_\mu$
discussed in section \ref{oneone}.
At the same time 
$U_1$ removes
the state 
$\left| 0,0 \right\ran$ from the theory.
This is very similar to the singular
gauge transformation in commutative
theory, but note that
MvN gauge transformation is concretely
defined and there appears no singularity.
Recall that the Wick
symbol of the 
projection operator
$\left| 0,0 \right\ran \left\lan 0,0 \right|$
concentrates around the origin of ${\R}^4$.
Since $\left| 0,0 \right\ran$ is removed from
$\H_{\cal I}$ by $U_1$, 
the Wick symbol of the
operator
$O^{(p)}$ acting in the
restricted subspace $\H_{\cal I}$
vanishes at the origin (see (\ref{van})):
\bea
\left.
\Omega_N(O^{(p)}) \right| 
_{(z_1,z_2)=(0,0)} =
\left\lan 0,0 \right| O^{(p)}
\left| 0,0 \right\ran  = 0.
\ena
This means that in the Wick symbol representation
the origin of ${\R}^4$ is removed
from the theory by MvN gauge transformation 
$U_1$.\footnote{This is
the reason why we have chosen the Wick symbol.
As we have stated in section \ref{opsymbol},
we regard operators as fundamental objects and
regard operator symbols as mere representations.
The advantage of the 
Wick symbol is that it gives clear topological picture
of projection operators.
}
As an illustration, we put the graph of
$x^1=x^3=0$ slice of the Wick symbol of 
$\left(\frac{\zeta}{4} \right)^2
\frac{1}{16}
F_{\mu\nu}^{(p_{\cal I})}F^{{(p_{\cal I})}\mu\nu}$,
see fig.\ref{ffone3D}.\footnote{Note that
this quantity is not gauge invariant.
}
It vanishes at the origin of ${\R}^4$
(for the calculation of this Wick symbol, 
see the end of this section).
Thus when represented in the
Wick symbols, MvN gauge transformation 
induces topology change: It extracts
a point from ${\R}^4$.
However this is only a matter of 
representation in operator symbols 
and at the operator level nothing essentially
changes because it is mere a change of the name.
However the topological 
features of gauge field configuration
is described only by using the operator symbols,
and when represented in the Wick symbol,
MvN gauge transformation gives
a unified 
description for the topology change
and gauge transformation.
This mixing of gauge field and
geometry is one of the most intriguing nature
of the gauge theory on noncommutative space.
From the IIB matrix model
points of view, gauge field
and geometry are indistinguishable:
In (\ref{bgf}) we regard $ i \hat{\pa}_\mu$ as
background and $i A_\mu$ as fluctuation,
but the original variable in
the IIB matrix model is $X_\mu$.

\subsubsection*{Calculation of 
$\left(\frac{\zeta}{4} \right)^2
\frac{1}{16}
F_{\mu\nu}^{(p_{\cal I})}F^{{(p_{\cal I})}\mu\nu}$
of $U(1)$ One-Instanton Solution}

The field strength in $\H_{\cal I}$
can be calculated from (\ref{PFS}):
\bea
F_{z_1\bar{z}_1}^{(p_{\cal I})} =
- F_{z_2\bar{z}_2}^{(p_{\cal I})}
&=&
\frac{\zeta}{\left(\frac{\zeta}{2}\right)^3 
\hat{N}(\hat{N}+1)(\hat{N}+2)} \, 
(z_1\bar{z}_1 - z_2\bar{z}_2), \nn
F_{z_1\bar{z}_2}^{(p_{\cal I})}
=F_{z_2\bar{z}_1}^{{(p_{\cal I})}\dagger}
&=& 
\frac{\zeta}{\left(\frac{\zeta}{2}\right)^3 
\hat{N}(\hat{N}+1)(\hat{N}+2)} \,
2 z_2\bar{z}_1,
\ena
where $\frac{\zeta}{2}\hat{N} = z_1\bar{z}_1
+ z_2 \bar{z}_2$.
The
instanton number is given by
\bea
\label{FtilF}
-\frac{1}{16\pi^2} 
\left(
2\pi \frac{\zeta}{4}
\right)^2
\mbox{Tr}_{\cal H}
F_{\mu\nu}^{(p_{\cal I})}\tilde{F}^{(p_{\cal I})\mu\nu}.
\ena
From (\ref{cpxF}) and (\ref{cpxF2}),
$
\frac{1}{16}F_{\mu\nu}^{(p_{\cal I})}\tilde{F}^{(p_{\cal I})\mu\nu}
=
F_{z_1\bar{z}_1}^{(p_{\cal I})}F_{z_2\bar{z}_2}^{(p_{\cal I})}
-
\frac{1}{2}
\left(
F_{z_1\bar{z}_2}^{(p_{\cal I})}F_{z_2\bar{z}_1}^{(p_{\cal I})}
+F_{z_2\bar{z}_1}^{(p_{\cal I})}F_{z_1\bar{z}_2}^{(p_{\cal I})}
\right)
$ and (\ref{FtilF}) becomes
\bea
& &\mbox{(\ref{FtilF})} \nn
&=&
-\frac{1}{\pi^2} 
\left(
2\pi \frac{\zeta}{4}
\right)^2
\mbox{Tr}_{\cal H}
\left(
F_{z_1\bar{z}_1}^{(p_{\cal I})}F_{z_2\bar{z}_2}^{(p_{\cal I})}
-
\frac{1}{2}
\left(
F_{z_1\bar{z}_2}^{(p_{\cal I})}F_{z_2\bar{z}_1}^{(p_{\cal I})}
+F_{z_2\bar{z}_1}^{(p_{\cal I})}F_{z_1\bar{z}_2}^{(p_{\cal I})}
 \right) 
\right)    \nn
&=&
\frac{1}{\pi^2} 
\left(
2\pi \frac{\zeta}{4}
\right)^2
\mbox{Tr}_{\cal H}
\left(
\frac{\zeta}{\left(\frac{\zeta}{2}\right)^3 
\hat{N}(\hat{N}+1)(\hat{N}+2)}
\right)^2
\left\{
(z_1\bar{z}_1 - 
z_2\bar{z}_2 )^2
+2 z_2 \bar{z}_1 z_1 \bar{z}_2 + 2 z_2 \bar{z}_1 z_1 \bar{z}_2
\right\} \nn
&=&
-
\frac{1}{\pi^2} 
\left(
2\pi \frac{\zeta}{4}
\right)^2
\mbox{Tr}_{\cal H}\, \, 
\zeta^2 
\left(
\frac{\zeta}{2}
\right)^{(-6+2)}
\frac{1}{\hat{N}^2(\hat{N}+1)^2
(\hat{N}+2)^2}(\hat{N}^2+2\hat{N}) \nn
&=&
4 \sum_{(n_1,n_2)\ne (0,0)}
\frac{1}{N(N+1)^2(N+2)} \nn
&=&
4 \sum_{N=1}^{\infty}
\frac{1}{N(N+1)(N+2)} \nn
&=& 1 .
\ena
Thus we have checked again that
the instanton number is one.
The Wick symbol of  
$\left(\frac{\zeta}{4} \right)^2
\frac{1}{16}F_{\mu\nu}^{(p_{\cal I})}F^{(p_{\cal I})\mu\nu}$
is given by
\bea
\label{nFF}
& &\Omega_N 
\left(
\left(\frac{\zeta}{4} \right)^2
\frac{1}{16}F_{\mu\nu}^{(p_{\cal I})}F^{(p_{\cal I})\mu\nu}
\right) \nn
&=&
\sum_{(n_1,n_2)\ne (0,0)}
\frac{1}{(n_1+n_2)(n_1+n_2+1)^2(n_1+n_2+2)}
\frac{r_1^{2n_1}}{n_1!}
\frac{r_2^{2n_2}}{n_2!}
e^{-r_1^2-r_2^2}.
\ena
where $r_1^2 = \frac{2}{\zeta} ((x^1)^2+(x^2)^2 )$ and
$r_2^2 = \frac{2}{\zeta} ((x^3)^2+(x^4)^2 )$.
We can rewrite (\ref{nFF}) as follows:
\bea
\label{FFr}
\sum_{(n_1,n_2)\ne (0,0)}\!\!\!\!\!\!& &\!\!\!\!\!\!
\frac{1}{(n_1+n_2)(n_1+n_2+1)^2(n_1+n_2+2)}
\, \, \frac{r_1^{2n_1}}{n_1!}
\, \frac{r_2^{2n_2}}{n_2!}\,
e^{-r_1^2-r_2^2} \nn
&=&
\sum_{N=1}^{\infty}
\frac{1}{N(N+1)^2(N+2)}
\sum_{n_1=0}^N \frac{1}{N!}\,
\frac{N!}{n_1! \, (N-n_1!)}\, 
r_1^{2n_1}r_2^{2(N-n_1)}e^{-r_1^2-r_2^2} \nn
&=&
\sum_{N=1}^{\infty}
\frac{1}{N(N+1)^2(N+2)}\,
\frac{r^{2N}}{N!}\, e^{-r^2},
\qquad r^2 = r_1^2 + r_2^2 \, \, .
\ena

\newpage
\section{Summary and Future Directions}\label{secsum}
\subsection*{Summary}

In the lectures
I have clarified the
topological
origin of the instanton number of
$U(1)$ gauge field on 
noncommutative ${\R}^4$.
The mechanism 
that gives instanton number
is quite impressive:
the projection operator
reduces the instanton number
to the winding number of gauge field
around $S^1$ at infinity.
I would like to emphasise again
that using operator symbols,
every topological nature
of the noncommutative instantons
can be described clearly.
The existence
of non-singular $U(1)$ instantons
is due to the intriguing 
nature of the noncommutativity,
but it is not mysterious.

Noncommutative $U(1)$ instanton solutions
with instanton number $k$ has one-to-one
correspondence with the
codimension $k$ ideal and
we can consider the projection to
the ideal states.
In order to clarify the physical
meaning of these projection operators,
I introduced the notion of the
gauge theory on restricted subspace
of the Fock space $\H$.
Generalized gauge equivalence relation
follows from 
partial isometry between 
two infinite dimensional
subspaces of the Fock space.
Two projections related
by partial isometry is called
Murray-von Neumann 
equivalent.
We determined 
transformation rule of gauge field under 
partial isometry map, and call
it Murray-von Neumann (MvN) gauge transformation.
This formalism 
is not the
noncommutative
Yang-Mills theory in the
usual sense, but
it is a natural extension.
Indeed, 
MvN gauge transformation
contains the noncommutative counterparts
of the singular gauge transformation in
commutative space.
From the IIB matrix model
viewpoints partial
isometry is mere a change of names of states
in the Fock space, and should not change physics.
By using this formalism
the instanton number
is explained as dimension of projection
which characterizes instanton solution.
This is an algebraic description of 
the instanton number, 
and the MvN gauge transformation relates
the algebraic description of 
the instanton number to the
topological one.

In the Wick symbol representation,
the MvN gauge transformation induces
a noncommutative analog of topology change.
At first sight the equivalence relation among
spaces with different topology is terrifying,
but at the operator level nothing essentially
changes under the MvN gauge transformation.
This unified description of gauge field and geometry
is one of the most intriguing features of the
gauge theory on noncommutative space,
and fits very naturally to the
framework of the IIB matrix model.

\subsection*{Indication to the
Ordinary Gauge Theory Description}

In \cite{SW} it was conjectured that
there is a one-to-one map between
ordinary description and
noncommutative description. 
Motivated by this conjecture
some solutions of ordinary
$U(1)$ gauge theory were
constructed
\cite{SW}\cite{st}\cite{tera}\cite{BN}.
However non of these solutions have
asymptotic behaviour of 
noncommutative $U(1)$ instanton
discussed in section \ref{oneone}.
Therefore if there
exists such map between
ordinary  and
noncommutative descriptions,
these solutions must (at least approximately) correspond to
the gauge field restricted in $\H_{\cal I}$
discussed in section \ref{padhm}.
In this case
there is a projection at the
core of instanton, and the covariant derivative
is appropriately modified.
This modification of the covariant derivative
can be interpreted as topology change
noncommutative space.
This suggests that 
in corresponding ordinary description, 
if it exists,
there must be a
topology change from ${\R}^4$.
Furthermore, the existence of
MvN gauge transformation in noncommutative
space suggests that 
this topology change in ordinary description
is (some kind of) gauge dependent notion, and
there is a explanation without topology
change.
In \cite{BN}, it was conjectured
that ordinary description requires
a modification of base space topology,
which has some similarities
with the gauge theory in restricted
Fock space.
If one seriously try to establish
explicit relation between
ordinary  and
noncommutative descriptions in the case of
instantons, 
the discussions in the lecture notes
must be taken into account.
I would like to mention that
in the case of monopoles and BIons,
there are little bit clearer descriptions for
the change of 
geometry in ordinary descriptions
\cite{hassan}\cite{mono}\cite{Bak}%
\cite{Mat}\cite{Mori}\cite{GN}.

\subsection*{D-brane Charge and 
Noncommutative Geometry}
In string theory, instantons
describe the bound states of
D$p$-branes and D$(p+4)$-branes 
\cite{SI}\cite{pinp4}.
In this regard 
the relation between projection operator
and D-brane charge was first pointed out
in \cite{mine},\footnote{The $U(1)$ one-instanton
solution given in \cite{NS} has already indicated
the relevance of the
projection operators in instanton solutions
on noncommutative
${\R}^4$.}
and the relations between projection
operator, partial isometry
and the topological charge in noncommutative
gauge theory
was first
clarified in \cite{mine2}.\footnote{As far as I know.
I think it is the first article that treats
these subjects in the recent context 
of string theory,
but my knowledge of mathematics is limited
and comments are welcomed.}
In the case of branes in the closed 
string vacuum,
K-theoretical classification of D-brane charge 
\cite{KMM}\cite{Kwitt}
in certain NS-NS B-field background was
recently investigated in
\cite{SFTwitt}\cite{HKLM}\cite{Mto}\cite{HM}
extending the techniques introduced in \cite{GMS}.

\subsubsection*{Surface Term and Trace}
In mathematical definition,
trace must satisfy the following
``trace property''
\bea
 \label{trace}
\mbox{Tr} AB = \mbox{Tr} BA.
\ena
Here $A$ and $B$ are elements of
some algebra and $\mbox{Tr}$
is a trace of that algebra.
However in the lecture notes, we have
encounter the quantity like
\bea
 \label{surface}
\mbox{Tr}_{\H} [ \hat{\pa}_\mu , K^{\mu}] \ne 0 .
\ena
This means the
trace $\mbox{Tr}_{\H}$ 
and unbounded operator $\hat{\pa}_\mu$
do not
satisfy trace property (\ref{trace}).
However, as we have seen,
such quantity is
important because it gives
topological charges.
Thus in order to obtain 
physically interesting results,
we should consider about loosening
the requirement (\ref{trace}).
In \cite{NCSFT} 
noncommutative geometric formulation
of string field theory was proposed.
In recent progress in string theory
it is becoming very plausible that
the R-R charge of D-branes
is classified in the stringy algebra
\cite{SFTwitt}\cite{OVK}.
In this case if we loosen
the requirement (\ref{trace}) and
D-brane will be represented by some
unbounded operator which gives surface 
terms.\footnote{The ``surface terms''
here are not restricted to the surfaces
in real space-time.
In noncommutative geometry, 
algebra itself
is one of the elements 
which describe geometry, and here we
mean surface in a
noncommutative ``space'' described by stringy
algebra. Of course concrete realization
of such an idea is left to the future works.
}
In string field theory the trace property
(\ref{trace}) is proved
under certain conditions,
but it may be useful to carefully
re-examine these conditions.
The deep meaning of
the tachyon condensation conjecture
\cite{Sen}
will be clarified by
establishing the topological/algebraic description
of the tachyon potential in string field theory.
The techniques developed here
may help the analysis of 
topological as well as algebraic
classification of D-brane R-R charges.

\begin{center}
{\large \bf Acknowledgements} 
\end{center}
The contents of the lecture notes 
based on talks
given at APCTP-Yonsei Summer Workshop on
Noncommutative Field Theories,
Tokyo University (Hongo)
and Tokyo Institute of Technology.
I thank the audiences 
for questions and comments.
I also thank participants of 
the workshop for many interesting discussions,
especially I am grateful to S. Terashima for 
explanations and discussions.
I would also like to thank
organizers of the workshop for warm
hospitality.

\newpage
\appendix

\section{The Case of %
$U(2)$ Instanton}\label{appU2}

In this section we construct
$U(2)$ one-instanton solution and observe the
difference from the $U(1)$ solution.
In this section we set $\zeta = 2$ for simplicity.
A solution to the 
ADHM equation (\ref{ADHMzeta}) is given by
\bea
B_1 = B_2 = 0, \quad
I = (\sqrt{\rho^2 + 2} \, \, \, \, 0\, ), \, \,
J^{\dagger} = (\, 0\, \, \,  \rho  \,)  .
\ena
Then the operator ${\cal D}_z$
becomes
\bea
{\cal D}_z 
=
\left(
\begin{array}{cccc}
-z_2 & -z_1 & \sqrt{\rho^2 + 2} & 0 \\
 \bar{z}_1 & -\bar{z}_2 & 0 & \rho
\end{array}
\right).
\ena
The operator zero-mode can be obtained as
\bea
\Psi_{0} = 
\left(
 \begin{array}{cc}
 & \\
 \Psi^{(1)}_{0} & \Psi^{(2)}_{0}\\
 &
 \end{array}
\right)      \, ,       
\, & &\Psi^{(1)}_{0}
=
\left(
\begin{array}{c}
\sqrt{\rho^2 + 2}\, \bz_2 \\
\sqrt{\rho^2 + 2}\,  \bz_1   \\
z_1\bz_1 + z_2 \bz_2  \\
 0
\end{array} 
\right) \frac{1}
{ \sqrt{\hN (\hat{N}+2 + \rho^2)}  }, \nn
& & \Psi^{(2)}_{0} = 
\left(
\begin{array}{c}
 -\rho z_1 \\
  \rho z_2 \\
    0      \\
z_1\bz_1 + z_2 \bz_2 +2
\end{array}
\right)  \frac{1}
{ \sqrt{ (\hN+ 2) (\hat{N}+2 + \rho^2)}  } ,\nn
{}
\label{SDpsiS}
\ena
where $\frac{1}{\sqrt{\hN} }$ is 
defined as
\bea
 \label{invN}
\frac{1}{\sqrt{\hN} }
=
\sum_{(n_1,n_2) \ne (0,0)}
\frac{1}{\sqrt{n_1 + n_2}} \,
| n_1,n_2 \ran\lan n_1,n_2| ,
\ena
i.e. when we consider the 
inverse of $\sqrt{\hN}$
we omit 
the kernel of $\sqrt{\hN}$, that is,
$|0,0\ran$, 
from the
Fock space. 
Hence $\frac{1}{\sqrt{\hN} }$ is a well defined operator.
When $\rho = 0$, the contribution of $\Psi_{0}^{(2)}$ 
to the field strength vanishes whereas
$\Psi_{0}^{(1)}$ reduces to the 
operator zero-mode (\ref{FPsi0}) in $U(1)$ one-instanton 
solution.
This operator zero-mode $\Psi_{0}$
is normalized as 
\bea
\Psi^{\dagger}_{0} \Psi_{0} = p, 
  \label{NpsiS}
\ena
where $p$ is a projection in the algebra of 
$2\times 2$ operator valued matrices:
\bea
p = 
\left(
\begin{array}{cc}
\mbox{Id}_{\cal H}- | 0,0 \ran\lan 0,0 | & 0 \\
 0 & \mbox{Id}_{\cal H}
\end{array}
\right).
\ena
Although in the case where the gauge group
is $U(2)$ the vector zero-modes have not been
classified at the moment,
we can directly check that the equation 
\bea
{\cal D}_z^{\dagger}
\frac{1}{ {\cal D}_z {\cal D}_z^{\dagger}  }
{\cal D}_z
=
1 - \Psi_{0} \Psi^{\dagger}_{0}  ,
\ena
holds.
Therefore the covariant derivative operator
\bea
\hat{D}_{\mu}^{(p)} &=& p\hat{\pa}_{\mu}p + A_\mu^{(p)}, \nn
& &A_{\mu} ^{(p)}
= \Psi_{0}^{\dagger} (\pa_\mu \Psi_{0} ) p,
\ena
gives anti-self-dual field strength.

Since the projection $p$ has infinite rank
as an operator, it is
Murray-von Neumann equivalent to the
identity operator $\mbox{Id}_{\H} \otimes \mbox{Id}_{2}$.
So let us MvN gauge transform
$p$ to $\mbox{Id}_{\H} \otimes \mbox{Id}_{2}$.
In order to do so 
one seeks for the operator $U$ which
satisfies
\bea
 \label{UUp}
U^{\dagger}U = p, \quad  
UU^{\dagger} = \mbox{Id}_{\H} \otimes \mbox{Id}_{2}.
\ena
Of course there are (infinitely) many 
gauge equivalent 
choices for such $U$.
In the case where the gauge group is $U(2)$,
there is a choice which has a physically interesting
interpretation.
Let us consider following operator $U$ 
which satisfies (\ref{UUp}) \cite{mine2}:
\bea
 \label{nsing}
U^{\dagger} = 
\left(
 \begin{array}{cc}
z_2 \frac{1}{\sqrt{\hN+1}}  & z_1  \frac{1}{\sqrt{\hN+1} } \\
- \bz_1 \frac{1}{\sqrt{\hN+1 } } & \bz_2 \frac{1}{\sqrt{\hN +1}  } 
 \end{array}
\right).
\ena
In large $r$ limit
the Wick symbol of this operator
becomes
\bea
U^{\dagger} \rightarrow 
g^{\dagger} \equiv
\frac{1}{r}
\left(
  \begin{array}{cc}
     z_2 & z_1 \\
     -\bz_1 & \bz_2
  \end{array}
\right)  \qquad (r \rightarrow \infty).
\ena
The asymptotic form of the gauge field
$A_\mu = (U\Psi_0^{\dagger}) \pa_\mu (\Psi_0 U) $
becomes
\bea
A_\mu   \rightarrow g \pa_\mu g^{\dagger} \quad 
(r \rightarrow \infty).
\ena
Thus in the case of $U(2)$ gauge theory
we can understand the topological
origin of the instanton number
in the same way as in the ordinary case.
Note that (\ref{nsing}) is a noncommutative
analog of the singular gauge transformation 
(see for example \cite{mine2}\cite{Raja}),
which is smoothened by the noncommutativity.

\newpage
\section{$U(1)$ Instantons on 
Noncommutative ${\R}^4$, Ideal
and Projections}\label{appNakaj}

\subsection{Ideal}\label{appideal}

Suppose we are given
a ring $R$. A subset ${\cal I} \in R$ is 
called {\bf ideal}
if it satisfies
following
two conditions:
\bea
\label{defideal}
a,b \in {\cal I}  \Longrightarrow a+b = {\cal I} ,\\
\forall r \in R, a \in {\cal I},
\, \, r \cdot a \in {\cal I}.
\ena
This definition of ideal
is taken from the 
property of the kernel of map. 
As an example,
let us consider
the ring of polynomials of complex number $z$
which we denote by ${\cal O}_z$.
Let us consider the subset $(z - w) {\cal O}_z$
which consists of all 
the polynomials of the form
$(z - w) f(z)$.
This subset is ideal since
\bea
(z-w) f(z) + (z-w) g(z) &=&  (z-w) (f(z) + g(z) ), \nn
r(z)\cdot (z-w) f(z) &=& (z-w)(r(z)f(z)).   \nn
& &\qquad 
f(z), g(z), r(z)\in {\cal O}_z.
\ena
This ideal is called 
`` ideal generated by $z-w$".
Let us consider the map $\phi_w$
from polynomials to complex number
${\cal O}_z \mapsto \C$ by
$\phi_w (f(z) ) = f(w)$.
Then $\ker \phi_w = (z - w) {\cal O}_z$,
that is, the ideal generated by $z-w$.

\subsection{One-To-One Correspondence Between \\
$U(1)$ Instantons on Noncommutative ${\R}^4$ and Ideal}

In this subsection
we review part of the beautiful
works of Nakajima \cite{Nakaj}\cite{LecNakaj},
which is relevant to our discussions:
We explain
one-to-one correspondence
between 
$U(1)$ $k$-instanton solution
 on noncommutative ${\R}^4$
and ideal ${\cal I} \subset {\cal O}_{{\C}^2}$  with
 $\mbox{dim}_{{\C}} \, {\cal O}_{{\C}^2}/{\cal I} = k$. 
First 
we show the isomorphism
\bea
 \label{isomor}
\ker {\cal D}_z
\simeq
{\cal I} =
\left\{
 f(z_1,z_2)
\Bigm| f(B_1,B_2) = 0
\right\}.
\ena
This isomorphism gives
a map from
the noncommutative
$U(1)$ instantons and
to the projections to the space of
ideal states $p_{\cal I}$.

Let us consider the space of
vector zero-modes
of ${\cal D}_z$,
$\ker{\cal D}_z =
\ker \tau_z \cap \ker \sigma^{\dagger}_z$:
\bea
\left| V \right\ran = 
\left(
 \begin{array}{c}
 \left| u_1 \right\ran \\
  \left| u_2 \right\ran \\
  \left|\, f \, \right\ran 
\end{array}
\right)  \quad \in 
\ker \tau_z \cap \ker \sigma^{\dagger}_z \quad .
\ena
The projection to third factor
\bea
\left| V \right\ran = 
\left(
 \begin{array}{c}
 \left| u_1 \right\ran \\
  \left| u_2 \right\ran \\
  \left|\, f \, \right\ran 
\end{array}
\right) \hookrightarrow \, 
\left|\, f \, \right\ran = 
f(z_1,z_2)   \left|\, 0,0\, \right\ran ,
\ena
gives the inclusion
\bea
\ker \tau_z \cap \ker \sigma^{\dagger}_z
\hookrightarrow {\cal H}.
\ena
We denote its image by 
${\cal H}_{\cal I}$.
It is easy to see
that for $f(z) \left|\, 0,0\, \right\ran 
\in {\cal H}_{\cal I}$,
\bea
& &f(z)\left|\, 0,0\, \right\ran 
   \in {\cal I} \nn
&\Leftrightarrow&
{}^{\exists}
u_1(z)\left|\, 0,0\, \right\ran ,
u_2(z)\left|\, 0,0\, \right\ran ,
\in {\cal H} \otimes {\C}^k \nn
& &\mbox{such that}
\quad 
f(z)\left|\, 0,0\, \right\ran I
=
(B_2-z_2)
u_1(z)\left|\, 0,0\, \right\ran 
+(B_1-z_1)
u_2(z)\left|\, 0,0\, \right\ran  \nn
&\Leftrightarrow&
f(B_1,B_2)I = 0.
\ena
Since the vectors of the form
$B_1^pB_2^q I, \, \,
p,q \in {\bf Z}_{\geq 0}
$ span
the whole $V = {\bf C}^k$ by 
the stability
discussed later in this appendix, it follows
that
$
f(z)\left|\, 0,0\, \right\ran
\in {\cal H}_{\cal I}
$
implies
$f(B_1,B_2)= 0$.

Conversely, suppose $f(B_1,B_2)= 0$.
Then, 
\bea 
 \label{CD}
& &f(z_1,z_2) \mbox{Id}_V   
= f(z_1-B_1+B_1,z_2-B_2+B_2)  \nn
&=&
f(B_1,B_2)
+
(z_2-B_2)(C(z) + (z_1-B_1)E(z) )
+
(z_1-B_1)(D(z) - (z_2-B_2)E(z) )     \nn
&=&
(z_2-B_2)(C(z) + (z_1-B_1)E(z) )
+
(z_1-B_1)(D(z) - (z_2-B_2)E(z) )  .
\ena
for some 
$C(z), D(z)$
and arbitrary 
$E(z)$ where
$C(z), D(z)$ and $E(z)$ are 
$k \times k$ matrices whose
entries are polynomials 
in $z_1$ and $z_2$.
We fix the ambiguity in $E(z)$ by requiring
\bea
(\bar{z}_1-B^{\dagger}_1)
C(z) 
-
(\bar{z}_2-B^{\dagger}_2)
D(z) = 0 .
\ena
This condition completely
eliminates ambiguity since
\bea
& &(\bar{z}_1-B^{\dagger}_1)
\tilde{C}(z) 
-
(\bar{z}_2-B^{\dagger}_2)
\tilde{D}(z) = 0               \nn
& & \mbox{for}\, \, 
\tilde{C}(z) = C(z) + (z_1-B_1)E(z) , \,
\tilde{D}(z) = D(z) - (z_2-B_2)E(z) \nn
&\Leftrightarrow &
(\bar{z}_1-B^{\dagger}_1) (z_1-B_1)E(z) 
+
(\bar{z}_2-B^{\dagger}_2) (z_2-B_2)E(z) = 0
\nn
&\Leftrightarrow &
(E(z) )^{\dagger} (\bar{z}_1-B^{\dagger}_1) 
(z_1-B_1)E(z) 
+
(E(z))^{\dagger} 
(\bar{z}_2-B^{\dagger}_2) (z_2-B_2)E(z) = 0
\nn
&\Leftrightarrow &
(z_1-B_1)E(z) = (z_2-B_2)E(z) = 0 \nn
&\Leftrightarrow &
\tilde{C}(z) = C(z),\quad 
\tilde{D}(z) = D(z).
\ena
If we write
$u_1(z) = C(z)I,
 u_2(z) = D(z)I$, we obtain
\EQ
f(z) I \left|\, 0,0\, \right\ran
=
-(B_2-z_2)
u_1(z) \left|\, 0,0\, \right\ran
-(B_1-z_1)
u_2(z)\left|\, 0,0\, \right\ran ,
\EN
with
\EQ
(B_1^{\dagger}-\bar{z_1})
u_1(z) \left|\, 0,0\, \right\ran
-
(B_2^{\dagger}-\bar{z_2})
u_2(z) \left|\, 0,0\, \right\ran = 0.
\EN
Therefore
$f(z)\left|\, 0,0\, \right\ran
\in {\cal H}_{\cal I}$.
This completes the proof
of (\ref{isomor}).

\subsubsection*{Stability}

The positivity of $\zeta$ ensures
the following stability \cite{Nakaj}\cite{nakatsu}
for any solution of ADHM equation
(\ref{ADHMzeta}):
\bea
 \label{stability}
&\bullet& 
\mbox{%
Any subspace $S$ of $V = {\bf C}^k$
which satisfies $B_a(S) \subset S \,
(a = 1,2)$ }\nn
& &\mbox{ and
Im $I \, \subset S$ is equal to $V$. }
\ena
Proof \\
Suppose (\ref{stability})
does not hold.
Then there exists subspace $S (\ne \emptyset)$
of $V$ such that $B_a(S) \subset S $
and Im $I\, \subset S$. Let $S_{\bot}$ 
be the subspace of $V$ orthogonal to
$S$.
Then $B_a$ and $B_a^{\dagger}$
acquire the form
\EQ
B_a = 
\left(
\begin{array}{cc}
 B_a|_S & D_a \\
  0     & (B_a^{\dagger}|_{S_\bot})^{\dagger}
\end{array}
\right)\quad
\left(
B_a^{\dagger} = 
 \left(
\begin{array}{cc}
 (B_a|_S)^{\dagger} & 0  \\
  D_a^{\dagger} & B_a^{\dagger}|_{S_\bot}
\end{array}
 \right)
\right)
\EN
Notice that the action of
$B_a$ and $B_a^{\dagger}$
are closed respectively
on $S$ and $S_\bot$.
$B_a|_S$ and 
$B_a^{\dagger}|_{S_\bot}$
are their restrictions on 
$S$ and $S_\bot$.
The restriction of  the
real ADHM equation (\ref{ADHMzeta})
to the subspace $S_\bot$
can be written as
\bea
\sum_{a}
[
(B_a^{\dagger}|_{S_\bot})^{\dagger},
 B_a^{\dagger}|_{S_\bot} ]
 - \sum_{a} D_a^{\dagger}D_a
- J|_{S_\bot}(J|_{S_\bot})^{\dagger}
= \zeta |_{S_\bot}
\ena
where
$J|_{S_\bot}$ is a projection of $J$
onto $S_\bot$.
Taking the trace of this equation
leads to a contradiction
because we set $\zeta > 0$.
Therefore $S = \emptyset$.
\vs{5}


From the stability condition (\ref{stability})
we can deduce dim $\H_{/{\cal I}} = k$.
We define a map $\phi$:
$\H \mapsto {\C}^k$
by 
$\phi \left( 
f(z_1,z_2) \left|0,0 \right\ran \right)
= f(B_1,B_2) I$.  
Since $B_\a (\mbox{Im}\, \phi) \subset \mbox{Im}\, \phi$  
($\a = 1,2$) and $\mbox{Im} \, \phi$ contains $I$, 
it must be
${\C}^k$ from the stability condition.
Hence $\phi$ is surjective.
We define ${\cal I} \equiv \ker \phi$.
Then ${\cal I}$ is an ideal in $\H$
and 
$\mbox{dim}\, \H_{/{\cal I}}=
\mbox{dim}\, (\H / \H_{\cal I}) = k$. 

Thus we have shown that
every noncommutative $U(1)$ instanton
has corresponding  
ideal ${\cal I} \subset {\cal O}_{{\C}^2}$ with
$\mbox{dim}_{{\C}} \, {\cal O}_{{\C}^2}/{\cal I} = k$.
The inverse is also true:
Every codimension 
$k$ ideal ${\cal I}$
has corresponding noncommutative 
$U(1)$ instanton with instanton number $k$.
To show this, we consider the
moduli space of $U(1)$
instantons on noncommutative ${\R}^4$. 
Recall that we can construct instantons
from the solution of the
ADHM equation (\ref{ADHMzeta}):
\bea
 \label{appADHMzeta}
\mu_{\R}
 &\equiv& [B_1 , B_1^{\dagger}] 
+  [B_2 , B_2^{\dagger}] 
        + II^{\dagger} - J^{\dagger} J 
  = \zeta \mbox{Id}_k ,\\
\mu_{\bf C} &\equiv& [B_1 , B_2 ] + IJ = 0 .
\ena
There is an action of $U(k)$ that does not change
the gauge field constructed by ADHM method:
\bea
 \label{appUk}
(B_1,B_2,I,J)
\mapsto
(u B_1 u^{-1}, u B_2 u^{-1} , u I, J u^{-1}),
\qquad u \in U(k).
\ena
Therefore the moduli space of
$U(1)$ instantons on noncommutative 
${\R}^4$
with instanton number $k$ is given by
\bea
  \label{moduli}
{\cal M}_{\zeta} (k,n=1)
 = 
\mu^{-1}_{\R}(\zeta) \cap \mu^{-1}_{\C}(0) /U(k) ,
\ena
where the action of $U(k)$ is the one
given in (\ref{appUk}). 
This moduli space is equivalently
described as follows \cite{LecNakaj}:
\bea
 \label{GS}
{\cal M}_{\zeta} (k,n=1)
\simeq \mu^{-1(s)}_{\C} (0) /GL(k),
\ena
where
\bea
\mu^{-1(s)}_{\C} (0)
\equiv
\Biggl\{
(B_\a , I, J) \Biggr|
\begin{array}{l}
(1) \, \mu_{\C} = 0 \\
(2) \, \mbox{stability condition} \, (\ref{stability})
\end{array} 
\Biggl.
\Biggr\} .
\ena
The proof is given in \cite{LecNakaj}.
Then we have the following isomorphism
\bea
 \label{IDEAL}
{\cal M}_{\zeta} (k,n=1)
\simeq
({\C}^2)^{[k]} \equiv
\Biggl\{ {\cal I} \subset {\cal O}_{{\C}^2} \Biggr| 
\begin{array}{l}
{\cal I} \, \mbox{is an ideal} \\
 \mbox{dim}\,  {\cal O}_{{\C}^2}/ {\cal I} = k
\end{array} 
\Biggl.
\Biggr\} .
\ena
Suppose we have have an ideal 
${\cal I} \in ({\C}^2)^{[k]}$.
Then we can construct a 
projection to the ideal states
$p_{\cal I}$.
Then we define 
$\H_{\cal I} \equiv p_{\cal I} \H$,
$ \H_{/{\cal I}}
\equiv
(1-p_{\cal I}) \H$ , and define 
$B_\a \in \mbox{End} \H_{/{\cal I}}$ as
the multiplication by $z_\a$ mod $\H_{\cal I}$
for $\a = 1,2$ and define 
$I \equiv \left| 0,0 \right\ran$ mod $\H_{\cal I}$.
It follows 
$[B_1, B_2] = 0$.
Since when $n=1$, one can show $J=0$
from $(1)$ and $(2)$ \cite{LecNakaj}, 
this condition is equivalent to
$\mu_{\C} = 0$.
The stability
condition holds
since $\left|0,0 \right\ran$
multiplied by  $z_\a$ span whole $\H$.
$GL(k)$ in (\ref{GS}) corresponds to the
change of basis (not orthonormal)
in $\H_{/{\cal I}}$.
\newpage
\section{Convention for the Complex Coordinates}
Complex coordinates:
\bea
z_1 = x^2 + i x^1 , \quad z_2 = x^4 + i x^3 .
\ena
Field strength:
\bea
F_{z_i\bar{z}_j} 
\equiv [D_{z_i},D_{\bar{z}_j}], \quad 
F_{z_iz_j} 
\equiv [D_{z_i}, D_{z_j}], \quad (i,j = 1,2)
\ena 
where
\bea
& &D_{z_1} = \frac{1}{2}(D_2 - iD_1),\quad
D_{z_2} = \frac{1}{2}(D_4 - iD_3), \nn
& &D_{\bar{z}_1} = \frac{1}{2}(D_2 + iD_1),\quad
D_{\bar{z}_2} = \frac{1}{2}(D_4 + iD_3).
\ena
Here $D_{\mu} = \pa_{\mu} + A_{\mu}$. Then we obtain
\bea
\label{cpxF}
& &F_{z_1\bar{z}_1} =\frac{1}{4} [(D_2 - iD_1),(D_2 + iD_1)]
= 
-\frac{i}{2}F_{12} , \nn
& &
F_{z_2\bar{z}_2} =\frac{1}{4} [(D_4 - iD_3),(D_4 + iD_3)]
= 
-\frac{i}{2}F_{34} ,\\
& &
F_{z_1z_2} 
= \frac{1}{4}[(D_2 - iD_1),(D_4 - iD_3)]
=\frac{1}{4}\left( F_{24}-F_{13} - i(F_{23}+F_{14}) \right),\nn
& &
F_{z_1\bar{z}_2} = \frac{1}{4}[(D_2 - iD_1),(D_4 + iD_3)]
=\frac{1}{4}
\left( F_{24}+F_{13} + i(F_{23}-F_{14}) \right). 
\label{cpxF2}
\ena
The anti-self-dual conditions: 
\bea
F_{z_1\bar{z}_1} + F_{z_2\bar{z}_2} = 0
&\Leftrightarrow &
F_{12} + F_{34} = 0 , \\
F_{z_1 z_2} =0
&\Leftrightarrow &
\left\{
\begin{array}{r}
F_{13} - F_{24}= 0, \\ 
F_{14} + F_{23}=0.
\end{array}
\right.
\ena
\newpage
\section{Some Formulas
in Operator Calculus}\label{oprel}
\begin{itemize}
\item Projection operators:
\bea
& &p^2 = p, \, \, p^{\dagger } =p ,\nn
& &p (1-p) = (1-p)p = 0.
\ena
Since for  
operators $A,B$ and $C$,
\bea
& &[A,BC] = [A,B]C + B[A,C], \nn
& &[AB,C] = [A,C]B + A[B,C],
\ena
we obtain following equations:
\bea
\label{oprel1}
p[\hat{\pa}_{\mu} , p]
&=& - p[\hat{\pa}_{\mu} , 1-p] 
=
- [\hat{\pa}_{\mu} , p(1-p)] 
+[\hat{\pa}_{\mu} , p] (1-p) \nn
&=&
[\hat{\pa}_{\mu} , p] (1-p), \nn
\quad [\hat{\pa}_{\mu} , p] p ,
&=& (1-p) [\hat{\pa}_{\mu} , p] \nn
p[\hat{\pa}_{\mu}, p] p &=& 0.
\ena

\item Similarity transformation:
\bea
\label{oprel2}
e^{A} \, B \, e^{-A}
=
B + [A,B] + \frac{1}{2!}[A,[A,B]] + \cdots 
=
\left[ 
\sum_{n=0}^{\infty}
\frac{1}{n!}
{\it \Delta}_A^n \right] \, B , 
\ena
where
${\it \Delta}_A B := [A,B]$.\\
\item When $[A,B]$ commutes with both $A$ and $B$,
\bea
 \label{oprel3}
e^A\, e^B &=& e^{[A,B]}\, e^B \, e^A \quad .\nn
e^A\, e^B &=& e^{\frac{1}{2}[A,B]}e^{A+B} \quad .
\ena

\end{itemize}
\newpage

\newpage
\newcommand{\NP}[1]{Nucl.\ Phys.\ {\bf #1}}
\newcommand{\AP}[1]{Ann.\ Phys.\ {\bf #1}}
\newcommand{\PL}[1]{Phys.\ Lett.\ {\bf #1}}
\newcommand{\CQG}[1]{Class. Quant. Gravity {\bf #1}}
\newcommand{\CMP}[1]{Comm.\ Math.\ Phys.\ {\bf #1}}
\newcommand{\PR}[1]{Phys.\ Rev.\ {\bf #1}}
\newcommand{\PRL}[1]{Phys.\ Rev.\ Lett.\ {\bf #1}}
\newcommand{\PRE}[1]{Phys.\ Rep.\ {\bf #1}}
\newcommand{\PTP}[1]{Prog.\ Theor.\ Phys.\ {\bf #1}}
\newcommand{\PTPS}[1]{Prog.\ Theor.\ Phys.\ Suppl.\ {\bf #1}}
\newcommand{\MPL}[1]{Mod.\ Phys.\ Lett.\ {\bf #1}}
\newcommand{\IJMP}[1]{Int.\ Jour.\ Mod.\ Phys.\ {\bf #1}}
\newcommand{\JHEP}[1]{JHEP\ {\bf #1}}
\newcommand{\JP}[1]{Jour.\ Phys.\ {\bf #1}}


\begin{thebibliography}{99}

\bibitem{CDS}
A. Connes, M. R. Douglas and A. Schwarz,
``Noncommutative Geometry and 
Matrix Theory: Compactification on Tori",
\JHEP{9802} (1998) 003.

\bibitem{NS}
N. Nekrasov and A. Schwarz,
``Instantons on Noncommutative ${\bf R}^4$ 
  and $(2,0)$ Superconformal Six Dimensional
  Theory", \CMP{198} (1998) 689.


\bibitem{mine}
K. Furuuchi,
``Instantons on Noncommutative
${\bf R}^4$ and Projection Operators",
Prog.\ Theor.\ Phys. {\bf 103} (2000) 1043,
hep-th/9912047.

\bibitem{mine2}
K. Furuuchi,
``Equivalence of Projections as Gauge Equivalence on
Noncommutative Space'',
hep-th/0005199.



\bibitem{ABS}
O. Aharony, M. Berkooz and N. Seiberg,
``Light-Cone Description of (2,0) 
Superconformal Theories in Six Dimensions",
Adv. Theor. Math. Phys. {\bf 2} (1998) 119.


\bibitem{SW}
N. Seiberg and E. Witten,
``String Theory and Noncommutative Geometry",
\JHEP{9909} (1999) 032.


\bibitem{IKKT}
N. Ishibashi, H. Kawai, Y. Kitazawa, A. Tsuchiya,
``A Large-N Reduced Model as Superstring",
\NP{B498} (1997) 467.

\bibitem{IKKTB}
H. Aoki, N. Ishibashi, S. Iso, H. Kawai, 
Y. Kitazawa, T. Tada
``Noncommutative Yang-Mills in IIB Matrix Model",
\NP{B565} (2000) 176.


\bibitem{Ho}
P.-M. Ho, 
``Twisted Bundle on Noncommutative Space
and $U(1)$ Instanton''
hep-th/003012.

\bibitem{Ours}
K. Furuuchi, H. Kunitomo and T. Nakatsu,
``Topological Field 
Theory and Second-Quantized Five-Branes",
\NP{B494} (1997) 144.

\bibitem{nakatsu}
T. Nakatsu,
``Toward Second-Quantization of D5-Brane",
\IJMP{A13} (1998) 923.

\bibitem{st}
M. Marino, R. Minasian, G. Moore, A. Strominger,
``Nonlinear Instantons from Supersymmetric p-Branes",
\JHEP{0001} (2000) 005.

\bibitem{tera}
S. Terashima,
``Instantons in the $U(1)$ Born-Infeld Theory 
and Noncommutative Gauge Theory''
Phys.Lett. B477 (2000) 292.

\bibitem{BN}
H. Braden and N. Nekrasov,
``Space-Time Foam From Non-Commutative 
Instantons",
hep-th/9912019.


\bibitem{hassan}
K. Hashimoto,
``Born-Infeld Dynamics in Uniform Electric Field''
\JHEP{9907}  (1999) 016.

\bibitem{mono}
A. Hashimoto and K. Hashimoto,
``Monopoles and Dyons in Non-Commutative Geometry'',
\JHEP{9911} (1999) 005.\\
K.~Hashimoto, H.~Hata and S.~Moriyama,
``Brane configuration from monopole 
solution in non-commutative super  Yang-Mills theory'',
JHEP {\bf 9912}, 021 (1999).\\
K.~Hashimoto and T.~Hirayama,
``Branes and BPS configurations 
of noncommutative / commutative gauge  theories'',
hep-th/0002090.

\bibitem{Bak}
D.~Bak,
``Deformed Nahm equation and 
a noncommutative BPS monopole'',
Phys.\ Lett.\  {\bf B471}, 149 (1999).

\bibitem{Mat}
D. Mateos,
``Noncommutative vs. commutative descriptions of D-brane BIons'',
\NP{B577} (2000) 139.



\bibitem{Mori}
S.~Moriyama,
``Noncommutative monopole from nonlinear monopole'',
Phys.\ Lett.\  {\bf B485}, (2000) 278.


\bibitem{GN}
D.~J.~Gross and N.~A.~Nekrasov,
``Monopoles and strings in noncommutative gauge theory'',
JHEP {\bf 0007}, 034 (2000).\\
D.~J.~Gross and N.~A.~Nekrasov,
``Dynamics of strings in noncommutative gauge theory'',
hep-th/0007204.



\bibitem{SI}
E. Witten,
``Small Instantons in String Theory",
\NP{B460} (1996) 541.

\bibitem{pinp4}
M. R. Douglas,
``Branes within Branes",
hep-th/9512077.


\bibitem{KMM}
R. Minasian and  G. Moore,
``K-theory and Ramond-Ramond charge''
\JHEP{9711} (1997) 002.

\bibitem{Kwitt}
E. Witten,
``D-Branes And K-Theory''
\JHEP{9812} (1998) 019.


\bibitem{NCSFT}
E. Witten,
``Non-commutative Geometry and String Theory''
\NP{B268} (1986) 253.

\bibitem{SFTwitt}
E. Witten,
`` Noncommutative Tachyons And String Field Theory''
hep-th/0006071.

\bibitem{HKLM}
J. Harvey, P. Kraus, F. Larsen, E. Martinec,
``D-branes and Strings as Non-commutative Solitons''
\JHEP{0007} (2000) 042.

\bibitem{Mto}
Y. Matsuo,
``Topological Charges of Noncommutative Soliton''
hep-th/0009002.

\bibitem{HM}
J. Harvey and G. Moore,
``Noncommutative Tachyons and K-Theory''
hep-th/0009030.

\bibitem{GMS}
R. Gopakumar, S. Minwalla, A. Strominger,
``Noncommutative Solitons''
\JHEP{0005} (2000) 020.


\bibitem{OVK}
E. Witten,
``Overview Of K-Theory Applied To Strings''
hep-th/0007175. 



\bibitem{Sen}
A. Sen,
``Tachyon Condensation on the Brane Antibrane System",
\JHEP{9808} (1998) 012.


\bibitem{Cole}
S. Coleman,
``Uses of Instantons",
in {\it Aspects of Symmetry}
Cambridge University Press (1985).

\bibitem{Raja}
R. Rajaraman,
``Solitons and Instantons",
North-Holland (1982).

\bibitem{ADHMconst}
M. Atiyah, N. Hitchin, V. Drinfeld and Y. Manin, 
``Construction of Instantons",  
\PL{65A} (1978) 185.\\
E. Corrigan and P. Goddard, 
``Construction of instanton monopole solutions
and reciprocity",
\AP{154} (1984) 253. \\
S. Donaldson, 
``Instantons and Geometric
Invariant Theory",
\CMP{93} (1984) 453. 


\bibitem{Nakaj}
H. Nakajima, 
``Heisenberg algebra and Hilbert schemes 
of points on projective surfaces",
Ann. of Math. {\bf 145}, (1997) 379, alg-geom/9507012.\\
H. Nakajima, 
``Instantons and affine Lie algebra",
Nucl. Phys. Proc. Suppl. {\bf 46} (1996) 154, 
alg-geom/9510003.

\bibitem{LecNakaj}
H. Nakajima, 
``Lectures on Hilbert scheme of points on surfaces",
AMS University Lecture Series vol. {\bf 18} (1999).

\bibitem{WO}
N.E. Wegge-Olsen,
``K-Theory and $C^*$-Algebras'',
Oxford University Press (1993).

\end{thebibliography}
\end{document}